\documentclass[journal]{IEEEtran}
\usepackage{cite}

\usepackage{graphicx}
\usepackage{booktabs}
\usepackage{multirow}
\usepackage{color}
\usepackage[table,xcdraw]{xcolor}

\usepackage{amsmath}
\usepackage{amsthm}
\usepackage{algorithm}
\usepackage{algorithmic}

\usepackage{array}

\ifCLASSOPTIONcompsoc
 \usepackage[caption=false,font=normalsize,labelfont=sf,textfont=sf]{subfig}
\else
 \usepackage[caption=false,font=footnotesize]{subfig}
\fi
\usepackage{fixltx2e}
\usepackage{dblfloatfix}

\ifCLASSOPTIONcaptionsoff
 \usepackage[nomarkers]{endfloat}
\let\MYoriglatexcaption\caption
\renewcommand{\caption}[2][\relax]{\MYoriglatexcaption[#2]{#2}}
\fi
\usepackage{url}

\begin{document}
%
\title{Production Scheduling Identification: An Inverse Optimization Approach for Industrial Load Modeling Using Smart Meter Data}
%
%
%

\author{Ruike Lyu,~\IEEEmembership{Graduate Student Member,~IEEE}, Hongye Guo$^*$,~\IEEEmembership{Member,~IEEE}, Qinghu Tang,~\IEEEmembership{Graduate Student Member,~IEEE}, and Qixin Chen,~\IEEEmembership{Senior Member,~IEEE}, Chongqing Kang,~\IEEEmembership{Fellow,~IEEE}

\thanks{This work was supported in part by the National Natural Science Foundation of China under Grant 52107102, and in part by the National Natural Science Foundation of China under Grant 52321004.

R. Lyu, H. Guo, Q. Tang, Q. Chen, and C. Kang are with the State Key Laboratory of Power System Operation and Control, Department of Electrical Engineering, Tsinghua University, Beijing 100084, China (Corresponding author: Hongye Guo. Email: hyguo@tsinghua.edu.cn).}} 
\maketitle

\begin{abstract}
To cost-effectively manage the supply-demand balance of the power system, the flexibility of industrial users could be harnessed through demand-side response. To minimize the negative impact on the production of industrial users during demand-side response, general-purpose models such as the state-task network (STN) are widely used to model the energy-consuming constraints of industrial production processes. However, the required model parameters cannot be set because the required data are privately owned by industrial users and are not directly available, hindering the accurate modeling of industrial loads. In this paper, we propose production scheduling identification (PSI), an inverse-optimization-based approach for industrial load modeling under incomplete information. In PSI, industrial users' smart meter data are used to identify production scheduling parameters, thus addressing the problem of accurate load modeling when private data are unavailable. We implemented PSI with a modified STN and proposed a practical algorithm to obtain an effective solution. Numerical tests showed that PSI can identify the model parameters of a steel powder plant and a cement plant with acceptable accuracy, using only 21 days of hourly smart meter data. Compared with accurate models established with direct access to private data, the modeling error does not exceed 8.5\% and 5.2\%, respectively.
\end{abstract}


\ifCLASSOPTIONpeerreview
\begin{center} \bfseries EDICS Category: 3-BBND \end{center}
\fi

%
\IEEEpeerreviewmaketitle
\section*{Nomenclature}
\addcontentsline{toc}{section}{Nomenclature}

\subsection*{Sets and Indices}
\begin{IEEEdescription}[\IEEEusemathlabelsep\IEEEsetlabelwidth{$I^{\rm P}/I^{\rm S}$}]
\item[$i/i^{\rm end}$] Index/total number of production tasks. $i=0$ for the feedstock.
\item[$I^{\rm P}/I^{\rm S}$] Set of production tasks/states. $I^{\rm P} = \{1, 2, ..., i^{\rm end}\}$, $I^{\rm S} = \{0\} \cup I^{\rm P}$
\item[$k/K$] Index/maximum number of iterations.
\item[$t/T$] Index/set of time intervals. $t=0$ for the initial time interval.
\item[$d$] Index/subscript of days.
\item[$m/L$] Index/library of production scheduling models.
\end{IEEEdescription}

\subsection*{The following are either parameters or variables, depending on the problem context:}
\begin{IEEEdescription}[\IEEEusemathlabelsep\IEEEsetlabelwidth{$S^{\rm 0/max/tar}_i$}]
\item[$Pr^{(d)}_{t}$] Electricity price at time interval $t$ (of day $d$).
\item[$E^{(d)}_{t}$] Metered electricity consumption in time interval $t$ of day $d$.
\item[$E_{dt}$] Modeled electricity consumption in time interval $t$ of day $d$.
\item[$Cost_d$] Energy cost on day $d$.
\item[$n$] Size of the dataset.
\item[$\Delta t$] Length of time intervals.
\item[$g_i$] Production in task $i$ achieved by consuming one unit of energy.
\item[$P_{dti}$] Operating power for task $i$ at time interval $t$ of day $d$.
\item[$P^{\rm max}_i$] Rated power of task $i$.
\item[$S_{dti}$] State (amount) of material $i$ at the end of time interval $t$ of day $d$.
\item[$S^{\rm 0/max/tar}_i$] Initial/upper limit/target state of material $i$.
\item[$\theta^{(d)}$]  Model parameter (updated for day $d$).
\item[$\mu^{\rm Star}_{di}$]  Dual variable for the production target constraints.
\item[$\mu^{\rm Smin/Smax}_{dti}$] Dual variable for the lower/upper state limit constraints.
\item[$\mu^{\rm Pmin/Pmax}_{dti}$] Dual variable for the lower/upper power limit constraints.
\item[$\lambda^{S}_{ti}$] Dual variable for the change of state constraints.
\item[$\alpha$] Adaptively decreasing parameter in model identification.

\end{IEEEdescription}

\section{Introduction}
%
%
%
%
\IEEEPARstart{I}{ncreasing} the share of renewable energy is paramount for addressing climate change but brings challenges to the supply-demand balance in global power systems~\cite{zhang_data-driven_2023}. In California, the prominent ``duck curve'' is steadily deepening~\cite{duck_curve}. Moreover, in Shandong Province, China, the frequent surplus of solar power at noon has led to a proportion of negative-price hours nearing 20\%~\cite{noauthor_-1h24____nodate}. To ensure power supply, redundant generation resources can be established, but this approach could prove costly~\cite{zhuo_cost_2022}. A more cost-effective solution lies in harnessing the flexibility of energy consumption among electricity users through grid-load interactions~\cite{wohlfarth_demand_2020}.
Among various electricity users, industrial users account for 50\% of global power consumption~\cite{wang_modelling_2019}. They are also more willing to respond to dynamic electricity prices by scheduling production in advance compared to residential consumers~\cite{wang_intelligent_2019}.
Therefore, to cost-effectively achieve carbon neutrality, addressing large-scale grid-industrial user interactions is an important step~\cite{chen_pathway_2021}.

To fully exploit the energy consumption flexibility of industrial users while minimizing negative impacts, it is necessary to establish accurate load models~\cite{cortez_demand_2023}. For industrial users, the energy consumed in the production process accounts for the vast majority of total energy consumption, making it a focal point in load modeling~\cite{golmohamadi_demand-side_2022}.
Following convention, we use the term ``load model'' to refer to the plant-level production scheduling model of industrial users, which delineates the constraints that need to be met when interacting with the grid~\cite{lyu_lstn_2023}. Specifically, the goal of modeling is to define the feasible region in which industrial users can schedule their production considering grid interactions at an hourly time scale, thereby preventing conflicts with industrial users' production targets or technical constraints during interactions.
In general, production processes in different industries may be subject to various types of technical constraints. For example, temperature variations in aluminum electrolysis or the iron molten process in steelmaking can be described via different high-precision mathematical models~\cite{niu_enhanced_2023}. However, in the context of grid-load interactions focused on hourly level energy usage, models that sacrifice some level of detail but offer greater generality and computational feasibility may be most suitable, such as the state-task network (STN)~\cite{kondili_general_1993, ding_demand_2014, lu_multi-agent_2020, lu_data-driven_2021, li_real-time_2017} and the resource-task network built on the basis of the STN~\cite{pantelides1994unified, zhang_cost-effective_2017, castro_resourcetask_2013}.

The widely used STN model represents materials as states and production processes as tasks, enabling a unified mathematical expression of the technical constraints in industrial production scheduling~\cite{kondili_general_1993}. Compared with load models that only reflect simple external characteristics, such as demand price elasticity~\cite{de_sa_ferreira_time--use_2013}, utility functions~\cite{maharjan_dependable_2013}, and cost functions~\cite{nojavan_optimal_2017} often used for residential loads, the STN model can accurately delineate the energy-material conversion relationships in production processes, the connections between production stages, and other constraints unique to industrial users. Therefore, the STN has been widely applied in the modeling of industrial loads. Ding et al.~\cite{ding_demand_2014} introduced a production scheduling scheme for oxygen generation systems utilizing the STN model, and processing tasks are arranged on the basis of day-ahead hourly electricity prices. Lu et al.~\cite{lu_data-driven_2021} proposed a production scheduling algorithm for a steel powder manufacturer based on the STN model, aiming to minimize electricity costs while meeting production requirements in response to real-time prices. The STN has also been applied to model typical industrial facilities such as cement plants~\cite{golmohamadi_robust_2020} and lithium-ion battery assembly systems~\cite{li_real-time_2017, lu_multi-agent_2020}.
In addition to applying the STN for load modeling across various industries, recent research has also delved into how to address uncertainties in factors such as electricity prices and orders~\cite{yu_real-time_2016, gong_integrated_2022}.
Yu et al. \cite{yu_real-time_2016} employed robust optimization to address future price uncertainties, developing a model that obtains real-time scheduling decisions for industrial loads while accounting for upcoming load commitments. Gong et al.~\cite{gong_integrated_2022} utilized multiobjective optimization to manage electricity price uncertainties and mitigate the default risk of delivery orders. Additionally, the electricity consumption strategies of specific industrial facilities concerning electricity prices and potential participation in spinning reserve markets have been studied~\cite{zhang_industrial_2015}.

While research on modeling and optimizing production scheduling for industrial users via load models such as the STN has been fairly comprehensive, it has been based on the assumption that the parameters used for modeling are readily available. This paper focuses on the issue of this assumption being invalid.
Specifically, the parameters required for modeling the loads of industrial users, such as the rated power of production equipment, material storage constraints between production stages, and production objectives~\cite{kondili_general_1993}, are likely highly sensitive data, potentially even considered proprietary information to competitors of the plants. Naturally, when industrial users optimize their own energy use through production scheduling models, these parameters can be deemed accessible. However, given transaction costs and technological thresholds, large-scale grid-load interactions may necessitate the aggregation of multiple electricity users through load aggregators or virtual power plants~\cite{chen_real-time_2024}. In such scenarios, owing to privacy concerns, these third-party external entities may struggle to access the parameters required for load modeling. In other words, while researchers have developed industrial load models such as the STN, the incomplete information in practice makes obtaining precise models from industrial users a potential challenge.

Our goal is to model industrial users using only easily obtainable external information, thereby eliminating the barriers of privacy protection that may limit accurate modeling or the detailed investigations of factories that were previously required by virtual power plant operators and other third parties. Considering the differences in production processes and equipment parameters across various industries and factories, achieving this goal is generally challenging. Nevertheless, the proliferation of smart meters has provided an opportunity to infer the internal parameters of users via external measurement data~\cite{wang_review_2019}. Virtual power plants or retailers often install one smart meter for each user to monitor and record their hourly electricity consumption over a period. In other words, smart meter data involve only aggregated information for the entire industrial facility (rather than individual devices) at a coarse temporal scale, which is generally considered to involve minimal privacy concerns but is indeed seen as providing valuable information for user modeling~\cite{wang_load_2015}. Therefore, in this paper, we explore the possibility of using smart meter data to indirectly acquire the internal parameters of industrial users required for load modeling.

Intuitively, the most relevant process of our goal is nonintrusive load monitoring (NILM), which involves analyzing the electricity meter data of users and disaggregating the data at the level of single internal pieces of equipment~\cite{tao_customer-centered_2023}. However, NILM for industrial users only provides load profiles for individual devices, and the parameters necessary for load modeling, such as material storage limits, cannot be appropriately set. Moreover, NILM typically requires high-frequency (e.g., 60 Hz) data collection~\cite{li_mixed-integer_2023}, exceeding the data resolution of smart meters. Owing to the limited information contained in hourly smart meter data, previous smart meter data analyses relied mainly on basic and general methods, such as classification and clustering~\cite{si_electric_2021, lu_electricity_2022, sun_clustering-based_2019}.
For example, Dehghan-dehnavi et al.~\cite{dehghan-dehnavi_estimating_2020} presented a two-level decision-making tree approach to estimate the DR potential of industrial facilities while considering various customer characteristics. To our knowledge, no prior research has addressed the challenge of accurate industrial load modeling using smart meter data.

Our work is inspired by the following idea: a rational industrial user would formulate a production scheduling optimization problem considering external electricity prices to minimize energy costs. Under this assumption, the hourly energy consumption recorded by smart meters can be considered the optimal energy usage on the basis of the user's load model. If an external entity knows the form of the user's load model but lacks certain parameters, these parameters can be inferred on the basis of the optimal energy usage results (smart meter data) and other problem parameters (electricity prices), constituting an inverse optimization problem by definition~\cite{ahuja_inverse_2001}. Inverse optimization has been applied in power system research~\cite{ruiz_2013_revealing, chen_2019_learning}, including load modeling for residential users~\cite{lu_2018_data}. Recent studies by Tan et al.~\cite{tan_data-driven_2023} indicated that compared with directly mapping boundary conditions to optimal solutions via machine learning models, inverse optimization-based methods exhibit higher accuracy because of the physics of the load models embedded in the constraints. Therefore, inverse optimization methods have the potential to aid in solving industrial load modeling problems under incomplete information conditions. However, the energy consumption dynamics of industrial users are more complex than those of residential users, necessitating further research into problem formulation and practical algorithms for identifying model parameters through inverse optimization.

In this paper, we introduce production scheduling identification (PSI), an inverse optimization-based framework for industrial load modeling under incomplete information conditions, utilizing historical smart meter data to identify the load model parameters of industrial users.
Specifically, we investigate the widely used STN as a general industrial load model and address the parameter identification issues for industrial users who are suitable for STN modeling.
To achieve the computationally tractable implementation of PSI, we modified the widely adopted STN model and developed an inverse optimization-based problem for load model parameter identification via the modified STN model (mSTN), along with a practical algorithm to solve it.
Numerical tests based on datasets from a steel powder manufacturer and a cement plant demonstrate that PSI with only 21 days of hourly metering data can be used to effectively identify the internal parameters of industrial user load models. The modeling error of load models obtained through PSI is within 8.5\% compared with directly acquiring true parameters. Our work validates the feasibility of industrial load parameter identification solely using smart meter data, offering a new solution for industrial load modeling under conditions of incomplete information.

Our contributions are therefore threefold:
\begin{itemize}
\item Our primary contribution lies in addressing the challenge of accurate load modeling for industrial users when internal data are unavailable. Existing research on industrial user modeling relies on internal parameters, which may not be directly accessible when privacy concerns arise. This issue is addressed in this paper through PSI, a framework that utilizes only the smart meter data of industrial users to identify their production scheduling parameters.
\item We have enhanced the STN model, a widely used industrial load general model known for its balance between complexity and representational capacity, to obtain the mSTN. The mSTN model has a linear form with per-unit values of model parameters and allows task aggregation, enabling the implementation of PSI to address parameter identification issues for industrial users that can be modeled via an STN.
\item We have introduced a practical algorithm to address the inverse optimization problem in PSI in a computationally feasible manner. A loss function has been designed to measure fitting errors, and data from multiple days are iteratively utilized. The numerical results validate the feasibility of our algorithm, demonstrating the viability of industrial load parameter identification solely through smart meter data.
\end{itemize}
The remainder of this paper is organized as follows: Section~\ref{sec_framework} presents the concept and framework of PSI. Section~\ref{sec_mSTN} describes the mSTN model, a computationally feasible industrial load model for implementing PSI. Section~\ref{sec_training} describes the model identification process of PSI. Sections~\ref{sec_case_study} and ~\ref{sec_discussion} present numerical results with realistic data and discussions. Section~\ref{sec_conclusion} contains the conclusions and prospects for future research.

\section{Concept and Framework of PSI}\label{sec_framework}
\begin{figure*}[!t]
  \centering
  \includegraphics[width=4.5in]{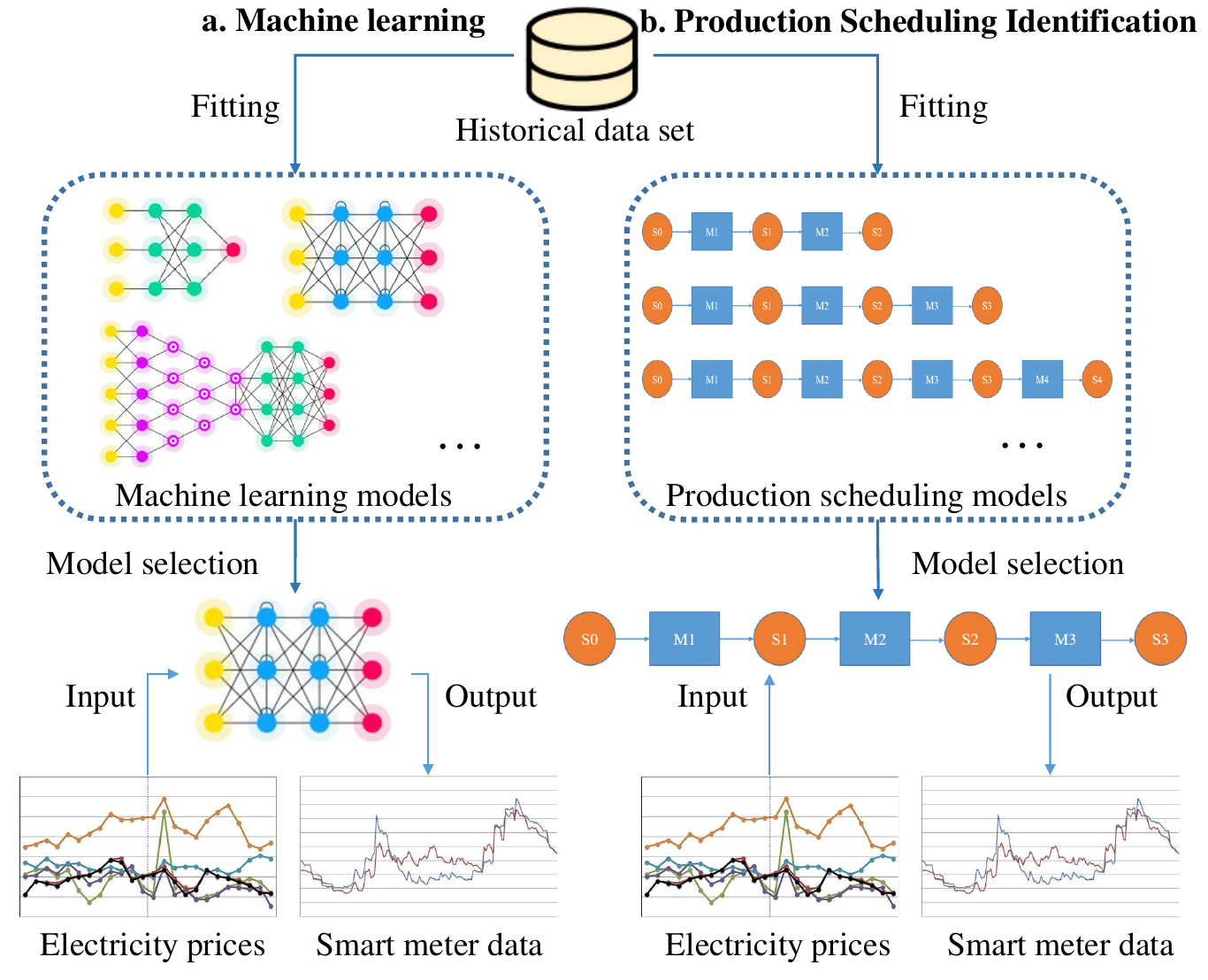}
\caption{Concept of production scheduling identification. The model identification process of PSI (b) is inspired by machine learning (a), but the model-free hypothesis in machine learning is replaced by the physics-based production scheduling problem in PSI, which is an optimization problem. With respect to the smart meter data of an industrial user, which are assumed to reflect optimal energy usage under electricity prices, the model identification process of PSI is formulated as an inverse optimization scheme, aiming to fit the parameters of an optimization problem on the basis of the solutions of the problem.}
  \label{fig_framework}
\end{figure*}
\subsection{Problem Description}

With the goal of establishing a load model when the internal parameters of industrial users are not accessible, this paper investigates how to use smart meter data to identify the parameters required for modeling. Specifically, we focus on modeling the industrial production process, where the necessary information for modeling can be categorized into specific facility parameters (private) and prior knowledge of the production process (public), explained as follows.

\paragraph*{Specific facility parameters} include internal information regarding the configuration of equipment, such as the rated power of equipment at each production stage, the material storage space, the production goals, and other parameters. These parameters involve the privacy of industrial users; therefore, without loss of generality, we consider them inaccessible directly.

\paragraph*{Prior knowledge of the production process} includes information that can be obtained through the Internet, expert knowledge, and other public channels unrelated to the specific parameters of equipment. In this work, we assume that the electricity and material consumption required for producing a unit product at each stage of production are known. For industries such as steelmaking, cement production, and aluminum electrolysis, which are typical energy-intensive industries, the processes are relatively mature, and the unit energy consumption at each stage can be obtained from national standards or industry reports. For example, a virtual power plant can easily determine that one aggregated factory is an aluminum electrolysis plant and can ascertain the quantities of electricity and alumina required to electrolyze 1 kg of aluminum without the need for onsite investigations or extra meter installations.

\paragraph*{Smart meter data} typically consist of average power readings taken every 15 minutes or hourly at various facilities. We assume that entities conducting parameter identification for retailers, virtual power plants, etc., have access to historical meter data (e.g., from the past month) for load modeling. Therefore, although these data are not publicly available, they are considered usable.

In addition to the above data, we consider the hourly electricity prices paid by industrial users to be known.
Formally, we use a historical price-consumption dataset $HD = \{(Pr^{(1)}, E^{(1)}), ..., (Pr ^ {(n)}, E ^ {(n)})\}$ to identify the load model parameters $\theta$, where $Pr^{(n)}$ and $E ^ {(n)}$ are the hourly electricity price and the user's hourly electricity consumption on day $n$, respectively; for the sake of conciseness, \(\theta\) encompasses the specific equipment parameters mentioned above and the prior knowledge of the production process.

\subsection{Overall Framework}

\paragraph*{Philosophy} PSI leverages the historical electricity consumption behaviors of industrial users and historical electricity prices to fit (train) a load model, which can be used to describe users' electricity-related behaviors under different boundary conditions. The idea of modeling user energy behavior on the basis of historical energy data is inspired by machine learning, for example, training a purely data-driven black-box model directly with electricity prices as inputs and smart meter data as outputs (Fig.~\ref{fig_framework}-a). However, this approach requires large quantities of high-quality data. The difference between the PSI framework and purely data-driven methods is that in PSI, a load model incorporating constraints from industrial production processes is used, thereby integrating the physics of industrial users' energy usage behaviors (Fig.~\ref{fig_framework}-b). Conceptually, PSI is a data-driven framework that incorporates domain knowledge, as the approach to modeling industrial users aligns with the actual physics of their electricity consumption processes, enhancing model effectiveness in scenarios with limited data. This characteristic mathematically and logically distinguishes PSI from purely data-driven methods and is key to achieving the accurate modeling of industrial loads via coarse-scale smart meter data.

\paragraph*{Assumptions} In this paper, the methodology of PSI is primarily based on the following two assumptions: first, we assume that an industrial user, faced with fluctuating electricity prices, optimizes their production schedule to minimize energy costs, which is a common assumption in research on interactions between industrial users and the grid. Second, although we do not specify the industry to which the industrial user belongs in the methodology section, we assume that the industrial user's load can be modeled via an STN. As mentioned earlier, production processes vary significantly across different industries, and even a generalized load model such as an STN cannot capture all the energy consumption characteristics of industrial users. Nevertheless, the STN model balances expressiveness and complexity, and it has been widely used to model cement plants, steel powder manufacturers, and lithium-ion battery assembly systems ~\cite{lu_data-driven_2021, golmohamadi_robust_2020, li_real-time_2017}. Therefore, we assume that the industrial user employs the STN to model their production processes and optimize their energy consumption accordingly.

\paragraph*{Methodology} Under the assumptions stated above, our objective is transformed as follows: if the smart meter data of an industrial user \(E ^ {(n)}\) under varying daily electricity prices \(Pr ^ {(n)}\) are the optimal energy usage results based on an STN model parameterized by \(\theta\), how can we determine \(\theta\)? The proposed inverse optimization-based method consists of three steps: first, enhancing the STN to derive the mSTN model (Section~\ref{sec_mSTN}), which uses the same load parameters as the STN but with a different mathematical form, aiming to obtain optimality conditions that enhance computational efficiency for solving inverse optimization; second, constructing an inverse optimization problem based on the mSTN, involving the design of a loss function to measure the fit of the results of a load model parameterized by \(\theta\) to the smart meter data (Section~\ref{sec_training}-A); and third, designing an algorithm to minimize the loss function with respect to \(\theta\) (Section~\ref{sec_training}-B\&C).

\begin{figure}[!t]
  \centering
  \includegraphics[width=3.0in]{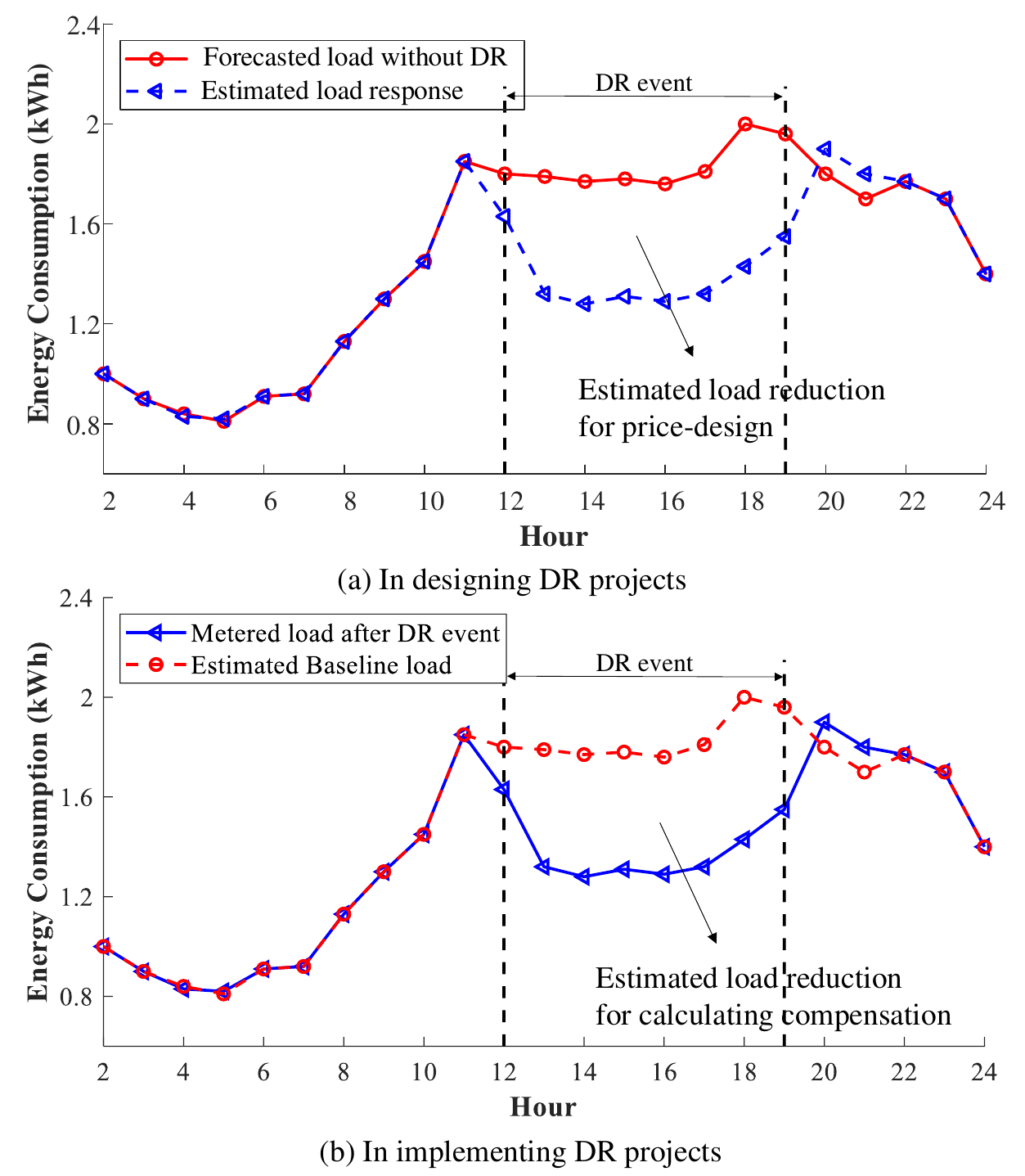}
\caption{Application examples of load modeling in DR projects. (a) In the design of DR projects: estimate the response of the electricity users under certain pricing schemes; (b) After DR events: estimate the baseline load (i.e., the load if the price remained unchanged and the DR event did not occur) for calculating the reduced load and the corresponding compensation for users.}
  \label{fig_illustration}
\end{figure}

\paragraph*{Applying identified load models} After the load parameters \(\theta\) of an industrial user are obtained in PSI, the load model is established in the form of an mSTN or STN. In general, the application scenarios of the obtained load model align with those of the widely used STN model, including estimating the demand response capacity of industrial facilities; additionally, the model is commonly embedded in economic dispatch or market clearing processes or integrated into the bidding models of load aggregators or virtual power plants~\cite{chen_real-time_2024}. As a simplified yet intuitive application, prior to initiating demand response actions, the hourly electricity consumption changes of users can be estimated by a virtual power plant when certain incentives are provided between 12:00 and 17:00 (Fig.~\ref{fig_illustration}(a)). Another application involves using the acquired load model for baseline load estimation~\cite{li_precision_2022}, calculating the original load profile of users without incentives, and enabling the computation of payments on the basis of a reduced load (Fig.~\ref{fig_illustration}(b)).

\section{Modified State-Task Network Model}\label{sec_mSTN}
The objective of this paper is to address the incomplete information problem in existing industrial load models applied in practical applications rather than improving the generality and accuracy of current models. Therefore, we chose the STN model~\cite{kondili_general_1993}, a general model widely used for simulating various industrial loads ~\cite{lu_multi-agent_2020}, and modified the STN model to enhance computational performance. Through linearization, the use of per-unit values, and aggregation, the mathematical form and scale of the subsequent inverse optimization problem were made tractable. This section presents the mSTN model that we obtained on the basis of modifications to the STN. Note that the PSI framework does not limit the form of the industrial load model.
For general-purpose load modeling, the mSTN can be used to implement PSI. Moreover, one could instead modify the load model or use other kinds of load models for specific cases and revise the optimality conditions in the model identification process (Section~\ref{sec_training}) accordingly.

\subsection{Modifications of the mSTN with Respect to the STN}

First, unlike the STN model, in which a machine can only operate in discrete states, the mSTN adopts a linearized form for two main reasons: a. For machines with very short start-up times, although their operating states are not continuous, the operating time within a time interval is close to continuous~\cite{zhang_demand_2018}, so using a continuous model is consistent with the actual process. b. The optimality conditions of linear programming are more computationally tractable with this approach, thereby improving the efficiency of the model identification process. The computational performance improvement and minimal increase in model error associated with the linearization of the STN model have already been studied, but there is a difference in the form of the model used here. Interested readers can refer to our previous work~\cite{lyu_lstn_2023}.

Second, to address the multivalue problem in fitting the load model, the parameters of the tasks are unitized.
The use of per-unit values is inspired by the fact that the state-related constraints are equivalent when $g_i$ and $S^{\rm 0/max/tar}_i$ are multiplied by a constant. This means that models with different parameters can be based on the same electricity-consuming strategy. Therefore, the mSTN uses per-unitized parameters in units of the final product (see Appendix~\ref{app_per_unitization}). For example, $S^{\rm 0}_0$ represents the amount of the final product that can be produced with the initial feedstock.

Third, the mSTN allows the aggregation of adjacent production tasks for two main reasons: a. In practice, the specific number of production tasks at an industrial facility may not be available for load modeling purposes, and the number of tasks, as a hyperparameter of the load model, can be adjusted by aggregating tasks, thus allowing the appropriate number of tasks to be determined through PSI. b. It can be theoretically shown that although aggregating tasks leads to reduced model expressiveness and possibly greater errors, such losses are acceptable (Appendix~\ref{app_aggregating}).
\subsection{Mathematical Formulation of the mSTN}

\begin{figure}[!t]
  \centering
  \includegraphics[width=3.0in]{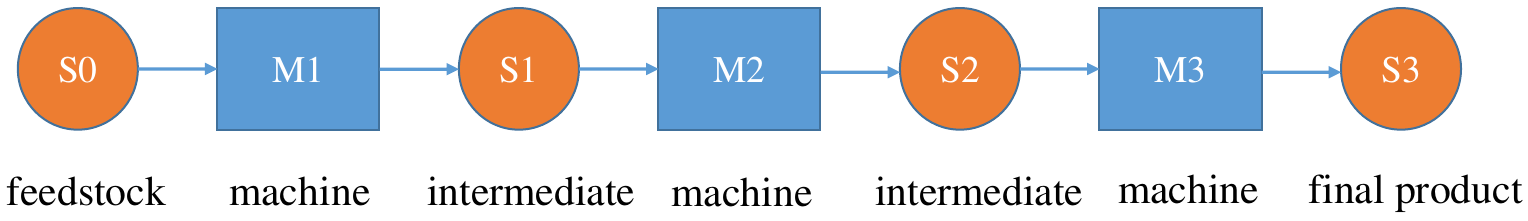}
\caption{A common state-task network.}
  \label{fig_mSTN}
\end{figure}

Fig.~\ref{fig_mSTN} shows a common industrial manufacturing process, where
${\rm M}_i$ and ${\rm S_i}$ represent the $i$th machine (``task'') and its product (``state''), both included explicitly as network nodes. The states can represent the feedstock, intermediate, or final products in the so-called task-state network. To maintain consistency with the symbol usage in the problem description section and to differentiate between actual data and model variables, in the following sections, we use superscripts in parentheses to represent the date on which the corresponding data were collected. For instance, \(E^{(d)}\) denotes a vector representing the energy consumption recorded by a smart meter on day \(d\), whereas \(E^{(d)}_t\) represents the element corresponding to period \(t\).
With respect to the variables in the load model, following convention, we uniformly use subscripts to denote entries. For example, \(E_{dt}\) represents the energy consumption of a facility in the mSTN on day \(d\) and in period \(t\).

On day $d$, a facility aims to minimize $Cost_d$, the total energy cost across the day, as formulated in (\ref{primal_cost}):
\begin{equation}\label{primal_cost}
  {\rm min.} Cost_d = \underset{t \in T}{\Sigma}{Pr^{(d)}_{t}} E_{dt},
\end{equation}
where the time intervals are set to be consistent with the electricity prices, e.g., hourly real-time prices.

Moreover, a facility must meet its production target (\ref{primal_constraint_tar}) and adhere to the technical constraints of the assembly line (\ref{primal_constraint_E})-(\ref{primal_constraint_storageLimit}). (\ref{primal_constraint_E}) shows the decomposition of the facility's hourly total electricity consumption for performing tasks.
(\ref{primal_constraint_Pmax}) ensures that the operating power required for each task does not exceed the rated power.
(\ref{primal_constraint_S0}) stipulates the initial states of the buffers.
(\ref{primal_constraint_changeofS1}), (\ref{primal_constraint_changeofS2}), and (\ref{primal_constraint_changeofS3}) represent the changes in buffer states over time for the feedstock, intermediate, and final products, respectively.
The buffer limit of the tasks is expressed by (\ref{primal_constraint_storageLimit}).
$\mu^{\rm Star}_{di}$,
$\mu^{\rm Smin/Smax}_{dti}$,
$\mu^{\rm Pmin/Pmax}_{dti}$, and
$\lambda^{S}_{ti}$ are the Lagrange multipliers of the corresponding constraints.
Following convention, we adopt Lagrange multipliers with the inequalities written as $\le 0$, although they are expressed in the above form for convenience.
\begin{flalign}\label{primal_constraint_tar}
  &\ S_{dti} \ge S^{0}_{i} + S^{\rm tar}_i \ : \mu^{\rm Star}_{di}, t = t^{\rm end}, i \in I^{\rm S}.
  &\\ \label{primal_constraint_E} %
  &\ E_{dt} = {(\underset{i \in I^{\rm P}}{\Sigma}P_{dti})}  \Delta t, t \in T.
  &\\ \label{primal_constraint_Pmax}%
  &\ 0 \le P_{dti} \le P^{\rm max}_i \
  : \mu^{\rm Pmin}_{dti}, \mu^{\rm Pmax}_{dti}, t \in T, i \in I^{\rm P}.
  &\\ \label{primal_constraint_S0}%
  &\ S_{dti} = S^{\rm 0}_i : \lambda^{S}_{dti}, \ i \in I^{\rm S}, t = 0.
  &\\ \label{primal_constraint_changeofS1}%
  &\ S_{dti} = S_{(t - 1)i}
  - P_{dt(i + 1)} g_{i + 1} \Delta t: \lambda^{S}_{dti}, t \in T, i = 0. 
  &\\ \label{primal_constraint_changeofS2}%
  &\  S_{dti} = S_{d(t - 1)i} + P_{dti} g_{i}  \Delta t
  - P_{dt(i + 1)} g_{i + 1}  \Delta t \\
  &\nonumber : \lambda^{S}_{dti}, t \in T, i \in I^{\rm P} \setminus \{i^{\rm end}\}. 
  &\\ \label{primal_constraint_changeofS3}%
  &\ S_{dti} = S_{d(t - 1)i} + P_{dti} g_{i} \Delta t
  : \lambda^{S}_{dti}, t \in T, i = i^{\rm end}.
  &\\ \label{primal_constraint_storageLimit}%
  &\ 0 \le S_{dti} \le S^{\rm max}_{i} \ : \mu^{\rm Smin}_{dti}, \mu^{\rm Smax}_{dti}, t \in T, i \in I^{\rm S}.&
\end{flalign}

In the mSTN, the decision variables for production scheduling are $\{P_{dti}, S_{dti}, E_{dt} | \forall i, \forall t\}$.
The problem parameters include external incentives $\{Pr^{(d)}_t|\forall t\}$ and the private data from the facility $\theta = \{g_i, P^{\rm max}_i, S^{\rm 0/max/tar}_i\}$.
As described above, within the PSI framework, the load model of industrial users represents a production scheduling optimization model. This means that users aim to minimize their energy costs under the production process constraints expressed via the mSTN.

\subsection{The dual problem of the mSTN}
The primary and dual problems in the mSTN are used to formulate the optimality conditions in the model identification process with PSI. As the mSTN model is a linear program, the dual problem can be conveniently written. The objective function is the Lagrangian dual function:
\begin{equation}\label{dual_g_function}
  \begin{aligned}
    {\rm max.} \ 
  & \underset{t \in T}{\Sigma}{Pr^{(d)}_{t}}
  {P_{t, 0}}\Delta t
  - \underset{t \in T}{\Sigma} \underset{i \in I^{\rm P}}{\Sigma} P^{\rm max}_i \mu^{\rm Pmax}_{dti} 
  - \underset{i \in I^{\rm S}}{\Sigma} S^{0}_{i} \lambda^{\rm S}_{d0i}\\
  &
  + \underset{t \in T}{\Sigma} \underset{i \in I^{\rm S}}{\Sigma} S^{\rm max}_i \mu^{\rm Smax}_{dti}
  + \underset{i \in I^{\rm S}}{\Sigma} (S^{0}_{i} + S^{\rm tar}_i)\mu^{\rm Star}_{di}
\end{aligned}
\end{equation}
The constraints include that the coefficients in the Lagrangian dual function equal 0:
\begin{flalign}\label{dual_constraint_coefficient}
  & Pr^{(d)}_{t} \Delta t - \mu^{\rm Pmax}_{dti} + \mu^{\rm Pmin}_{dti}
  + g_{i-1}\lambda^{S}_{dti-1} - g_{i}\lambda^{S}_{ti} = 0, \\
  &\nonumber \quad t \in T, i \in I^{\rm P}\\
  &\lambda^{S}_{ti} - \lambda^{S}_{d(t+1)i} - \mu^{\rm Smin}_{dti} + \mu^{\rm Smax}_{dti} = 0, \\
  &\nonumber \quad t \in T \setminus \{t^{\rm end}\}, i \in I^{\rm S}\\
  & \lambda^{S}_{ti} - \mu^{\rm Star}_{di} - \mu^{\rm Smin}_{dti} + \mu^{\rm Smax}_{dti} = 0,\\
  &\nonumber \quad  t = t^{\rm end}, i \in I^{\rm S}
\end{flalign}
The dual variables are nonnegative:
\begin{equation}\label{dual_constraint_non_negative}
  (\mu^{\rm Star}_{di}, \mu^{\rm Smin/Smax}_{dti}, \mu^{\rm Pmin/Pmax}_{dti}) \ge 0, \forall i, t \in T
\end{equation}

\section{Model Identification}\label{sec_training}

In this section, on the basis of the mSTN, historical price and consumption data for an industrial facility are used to identify a load model that can fit the facility's electricity consumption strategy.

\subsection{Problem Formulation}

The model identification problem based on inverse optimization can be expressed as (\ref{inverse_cost})-(\ref{inverse_dual_constraints}). (\ref{inverse_cost}) is a measure of the fitting error, which is the average squared residual of the load profile; $(\ref{primal_cost})=(\ref{dual_g_function})$ are strong duality conditions for the mSTN problem; $(\ref{primal_constraint_tar})-(\ref{primal_constraint_storageLimit})$ are primal constraints (except that the facility parameters now appear as variables); $(\ref{dual_constraint_coefficient})-(\ref{dual_constraint_non_negative})$ are dual constraints. (\ref{inverse_strong_duality}), (\ref{inverse_primal_constraints}) and (\ref{inverse_dual_constraints}) together formulate the optimality conditions of the mSTN model to avoid complementary relaxation constraints (nonlinear constraints) under Karush-Kuhn-Tucker conditions, which is a common approach to enhance the efficiency of solving inverse optimization problems~\cite{ruiz_2013_revealing}.
\begin{flalign}\label{inverse_cost}
&\  {\rm min.} \ \frac{1}{n} \sum_{d = 1}^{n} \sum_{t \in T} (E_{dt} - E^{(d)}_t)^2
&\\ \label{inverse_strong_duality}
&\  {\rm s.t.} \ {\rm strong \ duality:} (\ref{primal_cost})=(\ref{dual_g_function}), \forall d.
&\\ \label{inverse_primal_constraints}
&\  \quad \ \ {\rm primal \ feasibility:}(\ref{primal_constraint_tar})-(\ref{primal_constraint_storageLimit}), \forall d.
  &\\ \label{inverse_dual_constraints}
&\  \quad \ \ {\rm dual \ feasibility:} (\ref{dual_constraint_coefficient})-(\ref{dual_constraint_non_negative}), \forall d.&
\end{flalign}

Please note that in load model identification, the facility parameters are quantities that need to be determined via inverse optimization; therefore, they are decision variables rather than problem parameters. The variables of the model identification problem include the model parameters $\theta = \{g_i, P^{\rm max}_i, S^{\rm 0/max/tar}_i, \forall i\}$, the primary variables $\{P_{dti}, S_{dti}, E_{dt} | \forall i, \forall t\}$ and
the dual variables $\{\mu^{\rm Star}_{di}, \mu^{\rm Smin/Smax}_{dti}, \mu^{\rm Pmin/Pmax}_{dti}, \lambda^{S}_{ti}, \forall i, t \in T\}$. The problem parameters are electricity prices $\{Pr^{(d)}_t|\forall t\}$ and historical smart meter data $\{E^{(d)}_t| \forall t, \forall d\}$. Without loss of generality, these are the only data that external entities can obtain when privacy issues exist. Under these circumstances, variable product terms (quadratic form) exist under both sets of constraints (\ref{inverse_strong_duality}) and (\ref{dual_constraint_coefficient}), making the model identification problem an optimization problem with nonlinear constraints.

In practice, parameters such as the production goals of industrial users may change over time and can only remain consistent over a relatively short period. Therefore, the scenario of modeling industrial users is characterized by a small amount of usable data. Here, we implicitly assume that $\theta$ does not change over time, which is reasonable for short to medium time scales such as a month. The case in which the electricity consumption strategy varies over time is left for future work.

\subsection{Assumptions for Model Identification}\label{sec_assumption}

The above model identification problem is a large-scale nonlinear problem. Here, we present some assumptions to enhance the computational feasibility of the model identification process, none of which involve facilities' private data. Based on domain knowledge, one can revise the assumptions or introduce new assumptions for a specific plant.

\paragraph{Upper limit of value} To avoid numerical problems, the variable is restricted to no more than a large real number (e.g., 1e3):
\begin{equation}\label{assumption_a}
  \begin{aligned}
    &   (g_i, P^{\rm max}_i, S^{\rm 0/max}_i, \mu^{\rm Star}_{di}, \mu^{\rm Smin/Smax}_{dti}, \mu^{\rm Pmin/Pmax}_{dti}, \lambda^{S}_{ti}) \\
    & \le M, \ \forall i
\end{aligned}
\end{equation}
\paragraph{Change of state} After a day of production, the feedstock (i.e., raw materials, where $i=0$ represents the first production stage) should be reduced (\ref{asp_star0}), the amount of final product should be increased (\ref{asp_star2}), and the intermediates should remain at the same levels or be increased (\ref{asp_star1}):
\begin{flalign}\label{asp_star0}
&\  S^{\rm tar}_i \le 0, \ i = 0 
&\\ \label{asp_star1}
&\  S^{\rm tar}_i = 0, \ i \in I^{\rm P} \setminus \{i^{\rm end}\}.
&\\ \label{asp_star2}
&\  S^{\rm tar}_i \ge 0, \ i = i^{\rm end} &
\end{flalign}
\paragraph{Maximum power} In the historical operation of any facility, there has been at least one period in which all machines were operating at the rated power. Here, historical data refer to the data from the entire dataset instead of the data from a single day.
\begin{equation}\label{assumption_c}
  \underset{i \in I^{\rm P}}{\Sigma} P^{\rm max}_{i}  \Delta t = \underset{t}{\rm max} (E_t).
\end{equation}
\paragraph{Prior knowledge of the production process} Although private data from a facility are not available in most cases, we can leverage the type of facility and the production process to obtain some prior knowledge, which can be used to aid model identification. In the numerical test, we assume that $\{g_i, \forall i\}$ is known because the per unit value of $1/g_i$ is the electricity consumption of task $i$ when the facility produces one unit of the final product. Conceptually, $1/g_i$ is an intensive quantity that is related to only the process adopted by the facility instead of a private parameter associated with the facility. Therefore, we believe that it can be obtained by referring to relevant information.

Note that we have not assumed that the total number of tasks is known. The above assumption applies to an arbitrary model hypothesis, and the total number of tasks must be determined via model selection.

\subsection{An Iterative Model Identification Algorithm}

\begin{algorithm}[!t]
\caption{Model identification of PSI.}
  \label{alg_framework}
  \begin{algorithmic}[1]
    \renewcommand{\algorithmicrequire}{\textbf{Input:}}
    \REQUIRE
    historical electricity consumption and electricity prices pairs $\{(Pr^{(1)},E^{(1)}),...(Pr^{(n)},E^{(n)}) \}$, model library $L$, maximum number of iterations $K$.
    \renewcommand{\algorithmicrequire}{\textbf{Output:}}
    \REQUIRE a load model parameterized by $\theta$.
    \STATE partition the data set into a training set $TRAIN$ and a cross-validation set $CV$.
    \FOR{model $m \in L$ (number of tasks $i^{\rm end} = m$)}
      \STATE calculate $\theta^{(k)}$ via (\ref{alg_initialize}), $k=1,...,n$ .
      \STATE set $k = n + 1$, initialize $\theta = \frac{1}{n} \sum_{k=1}^{n} \theta^{(k)}$.
      \WHILE{$k \le K$}
        \STATE calculate $\theta^{(k)}$ via (\ref{alg_update})
        \STATE update $\theta = \theta^{(k-1)} + \frac{1}{n} (\theta^{(k)} - \theta^{(k-1)})$.
        \STATE \textbf{if} $\theta$ converges, \textbf{break}.
        \STATE set $k = k + 1$.
      \ENDWHILE
    \ENDFOR
    \STATE select model $m^*$ that performs best in $CV$.
  \end{algorithmic}
\end{algorithm}
The original form of the parameter identification problem is a large-scale optimization problem with nonlinear constraints, which is difficult to solve directly. Therefore, we draw inspiration from the zero-order stochastic gradient descent (ZO-SGD) approach commonly used in the machine learning field to solve similar problems~\cite{liu_primer_2020}. Here, stochastic gradient descent means that we do not use all the data at each iteration but randomly select a batch of data (defaulting to data from one day), thus reducing the scale of the problem at each iteration. Zero order means that we do not analytically compute the gradient but estimate it through a suboptimal search. The key steps of the proposed Algorithm~\ref{alg_framework} are explained as follows.

1-2: Without loss of generality, we use the training set $TRAIN$ to train each model $m$ in the model library $L$ and then use the cross-validation set $CV$ to select the best-performing model $m*$ as the output (step 12). The difference between the models lies in the number of aggregated tasks $i^{\rm end}$.

3-4: For initialization, for a given model $m$, we solve the following batch problems and calculate $\theta^{(k)}$, $k = 1,...,n$. In the $k$th calculation, following the convention of the stochastic gradient method, $d$ is randomly selected, and the following problem is solved:
\begin{flalign}
\label{alg_initialize}
  &{\rm min.} \sum_{t \in T} (E_{dt} - E^{(d)}_t)^2, \\ \nonumber
  &{\rm s.t.} (\ref{inverse_strong_duality}), (\ref{inverse_primal_constraints}), (\ref{inverse_dual_constraints}).&
\end{flalign}
to obtain $\theta^{(k)}$ and initialize $\theta^{\rm ref} = \frac{1}{n} \sum_{k=1}^{n} \theta^{(k)}$.

5-9: In the $k$-th iteration, $d$ is randomly selected, and the following equation set is solved:
\begin{flalign}\label{alg_update}
  &{\rm min.}  (\theta^{(d)} - \theta^{\rm ref})^2 + \alpha \sum_{t \in T} (E_{dt} - E^{(d)}_t)^2, \\ \nonumber
  &{\rm s.t.} (\ref{inverse_strong_duality}), (\ref{inverse_primal_constraints}), (\ref{inverse_dual_constraints}).&
\end{flalign}
to update $\theta^{(k)} = \theta^{(d)}$ and simultaneously update $\theta = \theta^{(k-1)} + \frac{1}{n} (\theta^{(k)} - \theta^{(k-1)})$. Compared with that in the parameter optimization problem in batch mode used during initialization (\ref{alg_initialize}), in the objective function in this case, $(\theta^{(k)} - \theta)^2$ represents the squared distance between the updated parameter and the reference parameter. Additionally, $\alpha$ adaptively decreases from 1 to 0 as $k$ increases to make the model identification process converge. Excluding these two modifications, at the $k$-th iteration, the objective is to first compute the optimal parameters \(\theta^{(k)}\) for the given batch data (\ref{alg_update}) and then approximate the gradient direction of the entire problem (\ref{inverse_cost}) in the direction of \(\theta^{(k)} - \theta^{(k-1)}\). Finally, \(\theta\) is updated with a fixed step size of \(\frac{1}{n}\).

Owing to the nonconvexity of the problem, it may take a long time to find the optimal solution at each iteration. For practical reasons, it is not necessary to find the optimal $\theta^{(d)}$ at each time step. Instead, a maximum computation time can be set for each iteration, and the obtained feasible solution when the maximum time is reached can be used to update the parameters. Commercial solvers can easily implement this strategy. As our core objective is to design an intuitive and practical algorithm to validate the feasibility of the PSI framework in principle, convergence analysis is omitted here owing to space constraints. Interested readers can refer to a related analysis of the ZO-SGD method~\cite{liu_primer_2020}.In addition, to accelerate the calculation process, the parameters corresponding to the data on different days can be updated in parallel. Algorithms to accelerate the PSI model identification process are left for future work.

\section{Numerical Results}\label{sec_case_study}

In this section, the feasibility and performance of PSI are verified using datasets from a cement plant and a steel powder manufacturer. Since no data-driven approach has been reported for modeling the load of industrial users, the methods we compare here are based on the primitive use of some mainstream machine-learning approaches. The detailed settings and codes can be found in ~\cite{rick10119_psi_2024}.

\subsection{Dataset and Partitioning}
\begin{figure*}[!t]
  \centering
  \includegraphics[width=7.0in]{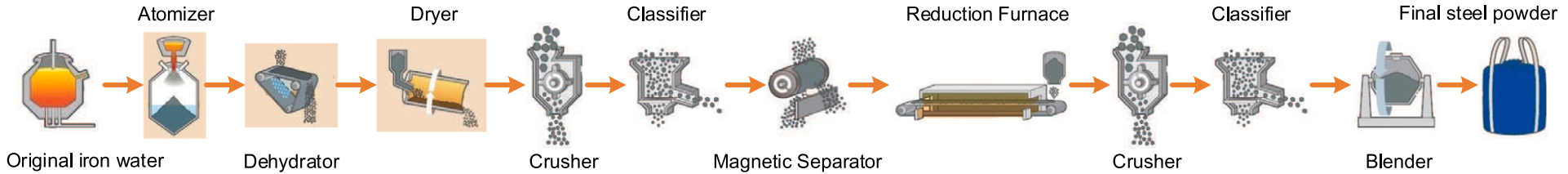}
\caption{The production process of a steel powder manufacturing facility.}
  \label{fig_process}
\end{figure*}

\begin{figure}[!t]
  \centering
  \includegraphics[width=3.0in]{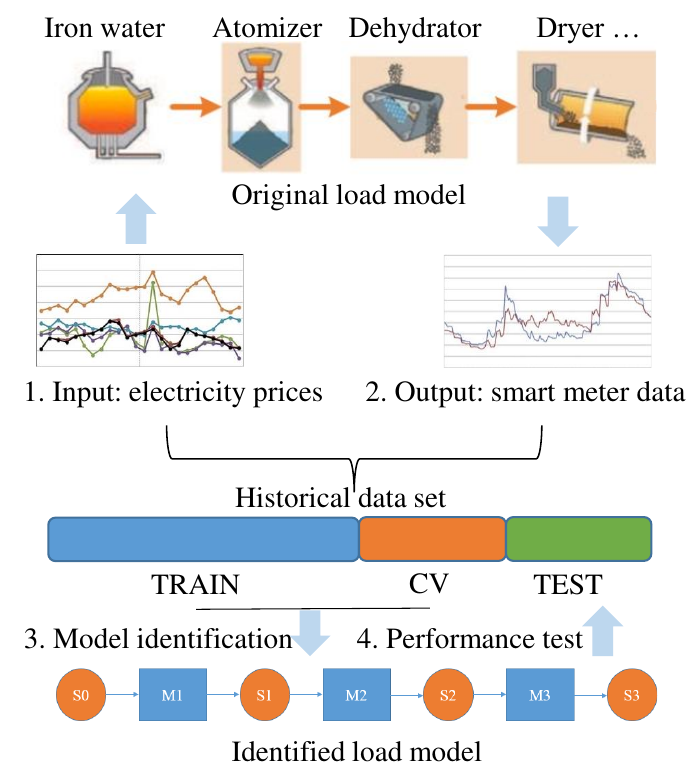}
\caption{Design of a numerical test. The electricity consumption data generated via realistic electricity prices and industrial facility parameters are used to establish a historical dataset, which is partitioned into a training set, a cross-validation set, and a test set.}
  \label{fig_case_study_design}
\end{figure}

\begin{figure}[!t]
  \centering
  \includegraphics[width=3.49in]{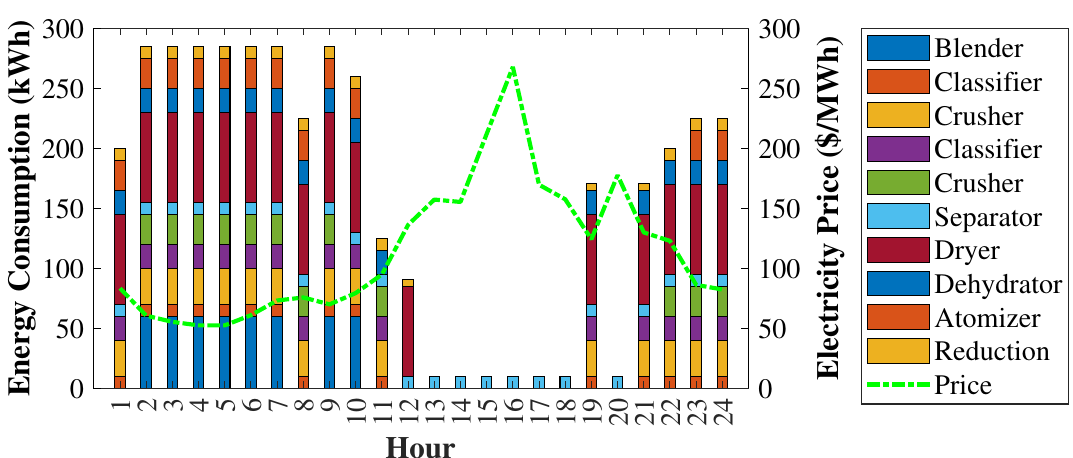}
\caption{Aggregated electricity consumption of all machines in the steel powder manufacturing system (Aug. 5th).}
  \label{fig_price}
\end{figure}

\begin{table}[!t]
  \caption{Parameters of the original steel powder manufacturing system.}
  \label{tab_parameter}
  \centering 
  \renewcommand{\arraystretch}{1.0}
  \begin{tabular}{lllll}
  \toprule
  \begin{tabular}[c]{@{}l@{}}Task name\\(number)\end{tabular} &
  \begin{tabular}[c]{@{}l@{}}
Operating
\\point\end{tabular} &
  \begin{tabular}[c]{@{}l@{}}Production\\rate\\(tons/h)\end{tabular} &
  \begin{tabular}[c]{@{}l@{}}Energy\\demand\\(kWh)\end{tabular} &
  \begin{tabular}[c]{@{}l@{}}Buffer\\capacity\\(tons)\end{tabular} \\ \midrule
                             & off  & 0  & 0  &                          \\ \cline{2-4}
\multirow{-2}{*}{Atomizer(1)}   & on & 30 & 60 & \multirow{-2}{*}{180}    \\ \hline
                             & off  & 0  & 0  &                          \\ \cline{2-4}
\multirow{-2}{*}{Dehydrator(2)} & on & 15 & 10 & \multirow{-2}{*}{100}    \\ \hline
\rowcolor[HTML]{FFFFFF} 
\cellcolor[HTML]{FFFFFF}     & off  & 0  & 0  & \cellcolor[HTML]{FFFFFF} \\ \cline{2-4}
\multirow{-2}{*}{\cellcolor[HTML]{FFFFFF}Dryer(3)} &
  on &
  15 &
  30 &
  \multirow{-2}{*}{\cellcolor[HTML]{FFFFFF}150} \\ \hline
                               & 1   & 0  & 0  &                          \\ \cline{2-4}
                             & 2   & 10 & 15 &                          \\ \cline{2-4}
\multirow{-3}{*}{Crusher(4,8)}    & 3   & 15 & 20 & \multirow{-3}{*}{100}    \\ \hline
                             & 1   & 0  & 0  &                          \\ \cline{2-4}
                             & 2   & 10 & 15 &                          \\ \cline{2-4}
\multirow{-3}{*}{Classifier(5,9)} & 3   & 20 & 25 & \multirow{-3}{*}{150}    \\ \hline
                             & off  & 0  & 0  &                          \\ \cline{2-4}
\multirow{-2}{*}{Separator(6)}  & on & 10 & 10 & \multirow{-2}{*}{100}    \\ \hline
 & off  & 0  & 0  &                          \\ \cline{2-4}
\multirow{-2}{*}{Reduction(7)}  & on & 15 & 75 & \multirow{-2}{*}{100}    \\ \hline
                             & 1   & 0  & 0  &                          \\ \cline{2-4}
                             & 2   & 10 & 6  &                          \\ \cline{2-4}
\multirow{-3}{*}{Blender(10)}    & 3   & 15 & 10 & \multirow{-3}{*}{200} \\  \bottomrule
\end{tabular}
\end{table}

\begin{table}[!t]
  \caption{Parameters of the original cement plant.}
  \label{tab_parameter_cement}
  \centering 
  \renewcommand{\arraystretch}{1.0}
  \begin{tabular}{lllll}
\toprule
\begin{tabular}[c]{@{}l@{}}Task\\ name\end{tabular}                               & \begin{tabular}[c]{@{}l@{}}
Operating
\\ point\end{tabular} & \begin{tabular}[c]{@{}l@{}}Production\\ rate (tons/h)\end{tabular} & \begin{tabular}[c]{@{}l@{}}Rated\\ power (kW)\end{tabular} & \begin{tabular}[c]{@{}l@{}}Buffer\\ capacity (tons)\end{tabular} \\ \midrule
\multirow{2}{*}{Crushing}                                                         & off                                                        & 0                                                                 & 0                                                          & \multirow{2}{*}{2000}                                           \\
                                                                                  & on                                                       & 1000                                                              & 2200                                                       &                                                                 \\ \hline
\multirow{2}{*}{\begin{tabular}[c]{@{}l@{}}Kiln feed \\ preparation\end{tabular}} & off                                                        & 0                                                                 & 0                                                          & \multirow{2}{*}{2500}                                           \\
                                                                                  & on                                                       & 250                                                               & 11000                                                      &                                                                 \\ \hline
\multirow{2}{*}{\begin{tabular}[c]{@{}l@{}}Clinker\\ production\end{tabular}}     & off                                                        & 0                                                                 & 0                                                          & \multirow{2}{*}{1750}                                           \\
                                                                                  & on                                                       & 300                                                               & 11550                                                      &                                                                 \\ \hline
\multirow{2}{*}{\begin{tabular}[c]{@{}l@{}}Finish\\ grinding\end{tabular}}        & off                                                        & 0                                                                 & 0                                                          & \multirow{2}{*}{5000}                                           \\
                                                                                  & on                                                       & 350                                                               & 11220                                                      &                                                                 \\ \bottomrule
\end{tabular}
\end{table}

We identified and tested the load model via daily price-smart meter data sequence pairs for 41 days (24 hours each day); these data were generated with realistic energy prices and industrial facility parameters (Fig.~\ref{fig_case_study_design}). The hourly real-time system energy prices in the PJM from 01.07.2022--10.08.2022 were used as the electricity prices. On the basis of the electricity price on each day, the original manufacturing model in MILP form was used to optimize the production scheduling of the facilities (Fig.~\ref{fig_process}) to generate hourly net electricity consumption for each day (Fig.~\ref{fig_price}). The equipment parameters of the original load model were taken from References~\cite{lu_data-driven_2021} and \cite{golmohamadi_robust_2020} and are listed in Tables~\ref{tab_parameter}\&\ref{tab_parameter_cement}, in which they are transformed into per-unit values. The production target for the final product is set at a maximum potential value, which corresponds to the output achieved when the slowest link in the production chain is continuously operated for 24 hours. The initial states of the buffers are set to half the buffer capacity. The data from the first 21 days are used as the training set ($n = 21$), the data from the middle 10 days are used as the cross-validation set to determine the hyperparameters of the model, and the data from the last 10 days (Aug. 1st to Aug. 10th) are used as the test set to assess model performance.

Note that we did not use the linear programming-based mSTN model to generate electricity consumption data for the facility but used the most representative STN model in the literature~\cite{lu_data-driven_2021} to generate data; this approach aided in verifying the generality of the proposed PSI framework and mSTN model. Additionally, the true value in the following results refers to the outcome of the STN model. We compared the mSTN model established with PSI through smart meter data with the STN model directly built with true parameters (which we consider the accurate model). Note that PSI does not require the user's energy consumption parameters, as modeling the user under the condition of unobtainable energy consumption parameters is the problem being solved with the PSI approach. Comparisons can be used to show whether our modeling method can achieve acceptable accuracy with incomplete information.

\subsection{Hardware, Schedule, and Task Descriptions}

We used Gurobi (V10.0.0) and MATLAB (R2021a) with YALMIP~\cite{Lofberg2004} to solve the optimization problems on a workstation with an Intel Core i9-10900X CPU (3.7 GHz) and 128 GB RAM. As mentioned before, for a given model hyperparameter, we assume that $g_i$, the produced material in stage $i$ using one unit of energy consumption, is known (Appendix~\ref{app_aggregating}), whereas the other parameters are undetermined. In the model identification process (Algorithm~\ref{alg_framework}), the maximum solution time in each iteration is set to 120 seconds. We ran the identification models for a maximum of 210 iterations, or 7 hours. The values of the variables after one iteration were used as the initial values of the variables for the next iteration.
We varied $\alpha$ in (\ref{alg_update}) throughout model identification, maintaining $\alpha = 1$ if $k \le 3n$ and consequently gradually decreasing it to 0.

After the identified load model was obtained, its performance was evaluated on the basis of the test set. The normalized RMSE (nRMSE) metric was used to evaluate the performance of the model:
$${\rm RMSE}=\sqrt{\frac{\sum^N_{i=1} (y^*_i - y_i)^2}{N}}$$
where $y^*_i$ and $y_i$ denote the hourly net energy consumption values for the trained model with identified parameters and the true values obtained with the STN model with true parameters, respectively. Normalization is achieved by dividing the absolute values by the maximum load.

\subsection{Results and Comparison}
\begin{figure}[!t]
  \centering
  \includegraphics[width=3.0in]{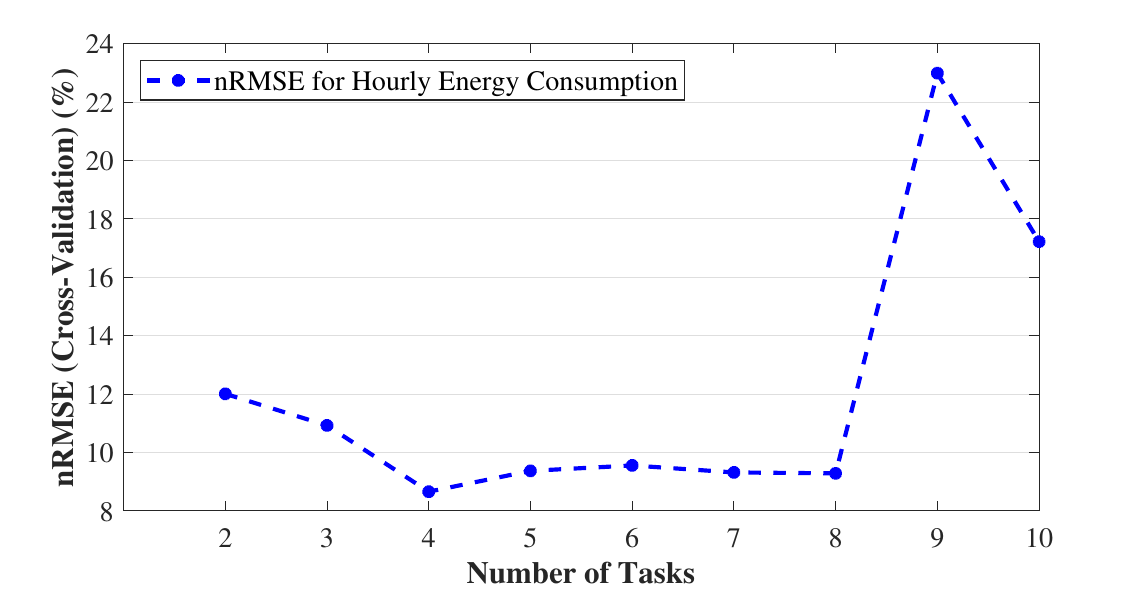}
\caption{Cross-validation performance of models of steel powder manufacturers with different hyperparameters.}
  \label{fig_rmse_wrt_m}
\end{figure}

\textbf{Model selection and identification results}. Fig.~\ref{fig_rmse_wrt_m} shows the value of the loss function in PSI for the cross-validation set when different hyperparameters of the mSTN model of a steel powder manufacturer are chosen. In general, as the number of tasks in the mSTN model increases, the performance of PSI first improves, achieving the best performance when the number of tasks is 4 or 5, and thereafter degrades. The numerical results imply that it is difficult to set sufficient parameter values when the number of tasks is large. This may be because the complexity of solving inverse optimization problems (\ref{alg_update}) increases significantly with increasing problem size. In practice, the number of tasks in the real process is likely to be unknown, and PSI can achieve good performance through model selection (even selecting a task number different from that of the real process), which verifies that task aggregation is reasonable. In the following tests, if not stated otherwise, the hyperparameter of the load model determined through model selection is $i^{\rm end} = 5$. Table ~\ref{tab_identified_model} lists the identified model parameters, where the task numbers with asterisks $*$ represent the production task numbers of the identified model, and the number in parentheses after is the original number of aggregated tasks; additionally, max. buffer refers to the difference between the buffer capacity and the initial state of the buffer, which is related to the number of intermediates that can be produced in advance. Intuitively, PSI can be used to identify most parameters of the load model with high accuracy. Owing to the relatively simple production process of the cement plant, which consists of only four main production stages, there is no need for further aggregation of the stages; therefore, no relevant results are presented here.

\begin{table}[!t]
  \caption{Identified Load Model Parameters for the Steel Powder Manufacturer.}
  \label{tab_identified_model}
  \centering 
  \renewcommand{\arraystretch}{1.0}
\begin{tabular}{llllll}
\toprule
\begin{tabular}[c]{@{}l@{}}Aggregated \\ task number\end{tabular} &
  \begin{tabular}[c]{@{}l@{}}Load model\\ parameter\end{tabular} &
  \begin{tabular}[c]{@{}l@{}}True \\ value\end{tabular} &
  \begin{tabular}[c]{@{}l@{}}Default \\ PSI\end{tabular} &
  \begin{tabular}[c]{@{}l@{}}No Assu-\\ mption b\end{tabular} &
  \begin{tabular}[c]{@{}l@{}}No Assu-\\ mption c\end{tabular} \\ \midrule
\multirow{2}{*}{$1^*$(1)}      & max. Power  & 60  & 56  & 44  & 39  \\ \cline{2-6} 
                        & max. Buffer & 90  & 80  & 106 & 36  \\ \hline
\multirow{2}{*}{$2^*$(2+3)}    & max. Power  & 40  & 37  & 55  & 272 \\ \cline{2-6} 
                        & max. Buffer & 75  & 67  & 320 & 56  \\ \hline
\multirow{2}{*}{$3^*$(4+5+6)}  & max. Power  & 55  & 54  & 55  & 53  \\ \cline{2-6} 
                        & max. Buffer & 50  & 8   & 301 & 8   \\ \hline
\multirow{2}{*}{$4^*$(7)}      & max. Power  & 75  & 80  & 69  & 85  \\ \cline{2-6} 
                        & max. Buffer & 50  & 72  & 199 & 264 \\ \hline
\multirow{2}{*}{$5^*$(8+9+10)} & max. Power  & 36  & 58  & 63  & 40  \\ \cline{2-6} 
                        & Target      & 240 & 222 & 85  & 218 \\ \bottomrule
\end{tabular}
\end{table}

\begin{table}[!t]
  \caption{The load model accuracy (nRMSE) for the two tested datasets.}
  \label{tab_rmse_methods}
  \centering
  \begin{tabular}{ccc}
    \toprule
    Approaches &
    \begin{tabular}[c]{@{}c@{}} Cement\\  plant\end{tabular} &
    \begin{tabular}[c]{@{}c@{}} Steel powder \\  manufacturer \end{tabular} \\ \midrule
      MLP  & 13.43\% & 18.50\% \\
  LSTM & 13.94\% & 19.20\% \\
  SVR  & 13.67\% & 19.11\% \\
  \textbf{PSI}  & \textbf{5.18\%}   & \textbf{8.52\%}  \\ \bottomrule
  \end{tabular}
  \end{table}

\begin{figure}[!t]
  \centering
  \includegraphics[width=2.0in]{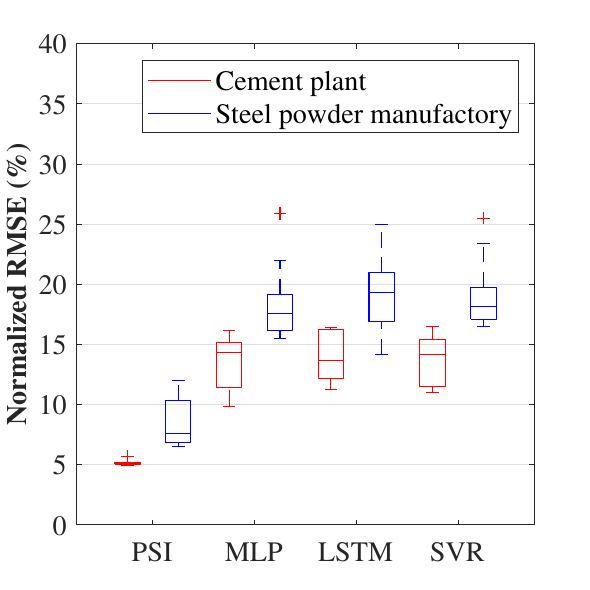}
\caption{Comparison of load modeling accuracy.}
  \label{fig_accuracy}
\end{figure}

\textbf{Performance of the identified load model}. In load modeling tasks with the test set, the PSI method outperforms mainstream machine learning approaches (multilayer perceptron (MLP), long short-term memory (LSTM), and support vector regression (SVR) methods). The detailed configurations are provided in Appendix~\ref{app_ml_methods}. Table~\ref{tab_rmse_methods} lists the accuracies of the compared methods. The nRMSEs of PSI for the two datasets (5.2\% and 8.5\%, respectively) are the lowest among those of the compared methods. The boxplot in Fig.~\ref{fig_accuracy} provides an intuitive representation of the modeling errors associated with different methods. The consistency in performance for the two different datasets validates the applicability of the PSI method, yet differences in results across the datasets also exist. For the cement plant dataset, the algorithm converged after 162 iterations, whereas for the steel powder manufacturing dataset, the iterative algorithm reached the maximum number of iterations at 210. Furthermore, for the cement plant dataset, the PSI results were better than those for the steel powder manufacturing dataset, which may be attributed to the fact that the cement production process involves only four stages, whereas steel powder manufacturing spans 10 stages. Therefore, the inverse optimization problem of identifying load parameters in the cement plant production process requires fewer decision variables and has a lower dimension, making it easier for the model to converge to a good result.

\textbf{Comparison with machine learning models}. The tested machine-learning models all produce worse estimates of electricity consumption than the proposed model. The primary reason is related to the characteristics of the test scenarios; namely, only small amounts of external meter data are available. The training set covers only 21 days, which is far less than the data period usually required for machine learning, resulting in underfitting. Nevertheless, this amount of data is consistent with the load modeling context in practice, especially considering that an industrial facility's electricity consumption is influenced mainly by the product orders received, and the electricity-consuming strategy of the facility will remain unchanged for only a short period, such as a few weeks. Moreover, in our settings, the factories can shift their energy usage with significant flexibility. This implies that the fluctuation in the load in the dataset is much greater than that of typical residential loads. Under these conditions, the difficulty of obtaining an accurate load model is far greater than that in traditional load forecasting.

\begin{figure}[!t]
  \centering
  \includegraphics[width=3.0in]{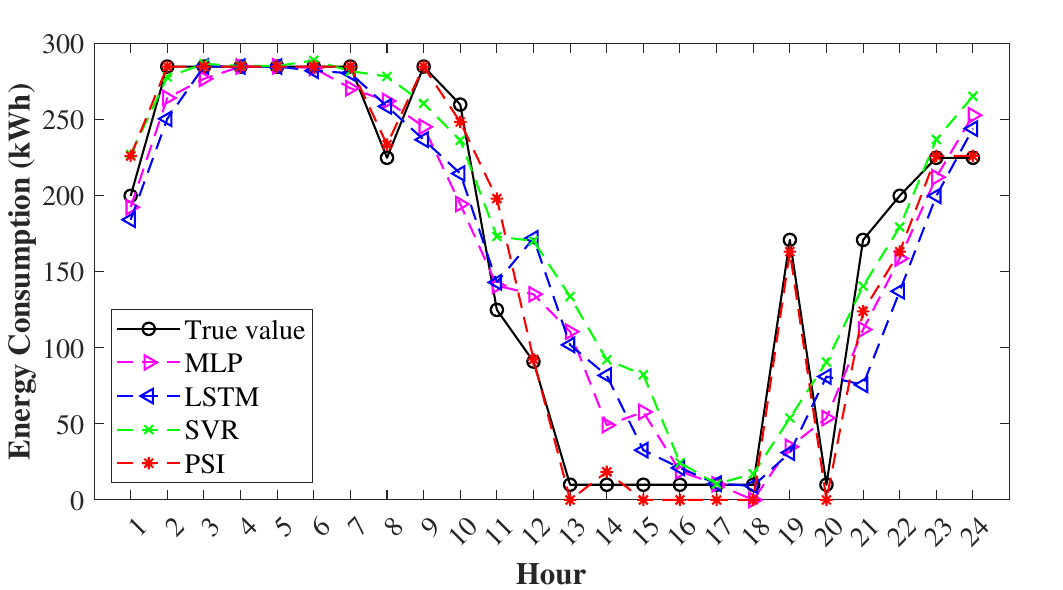}
\caption{Estimated load response of the steel powder manufacturer considering electricity prices (Aug. 5th).}
  \label{fig_typical_load_steel}
\end{figure}

\begin{figure}[!t]
  \centering
  \includegraphics[width=3.0in]{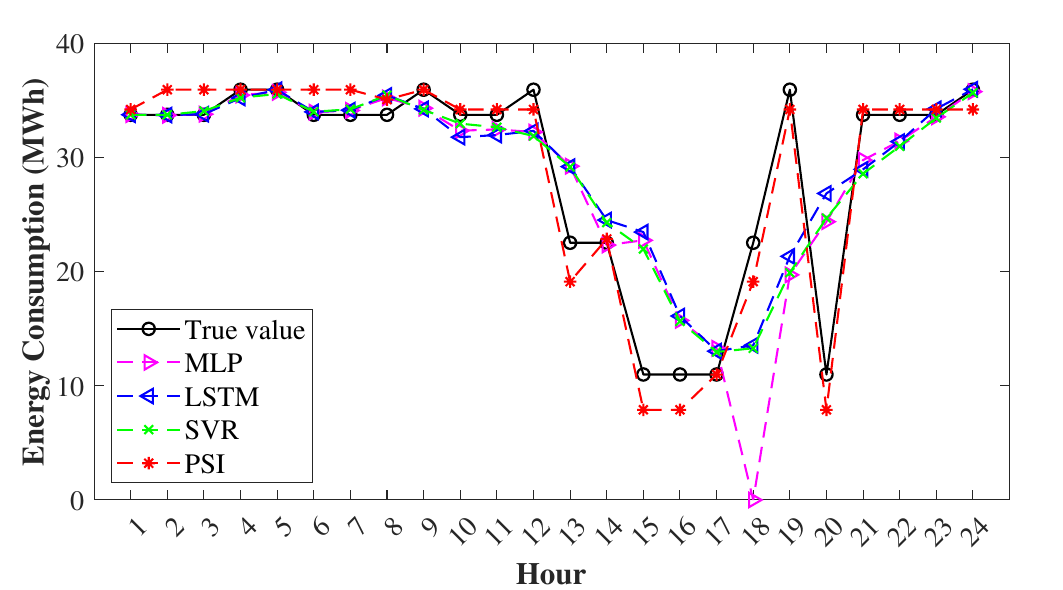}
\caption{Estimated load response of the cement plant considering electricity prices (Aug. 5th).}
  \label{fig_typical_load_cement}
\end{figure}

To illustrate the accuracy of the PSI method in load modeling tasks, Fig.~\ref{fig_typical_load_steel} and Fig.~\ref{fig_typical_load_cement} were established; they show the typical load profile and the output results given by the trained load models. An interesting observation is that the results of PSI are more accurate during high-electricity-price periods (e.g., 3:00 PM to 5:00 PM) than are the results of machine learning. This could be attributed to the assumption embedded in the PSI framework that users will optimize their energy costs; therefore, the energy usage results given by the load model established with PSI will avoid energy use in high-price hours to reduce costs. This is consistent with the actual energy usage behavior of users, as reflected in the lower RMSE. In comparison, the machine-learning models fail to fully learn the behavior characteristics of users avoiding usage in high-electricity-price periods, so their performance is worse at these times.

\begin{figure}[!t]
  \centering
  \includegraphics[width=3.0in]{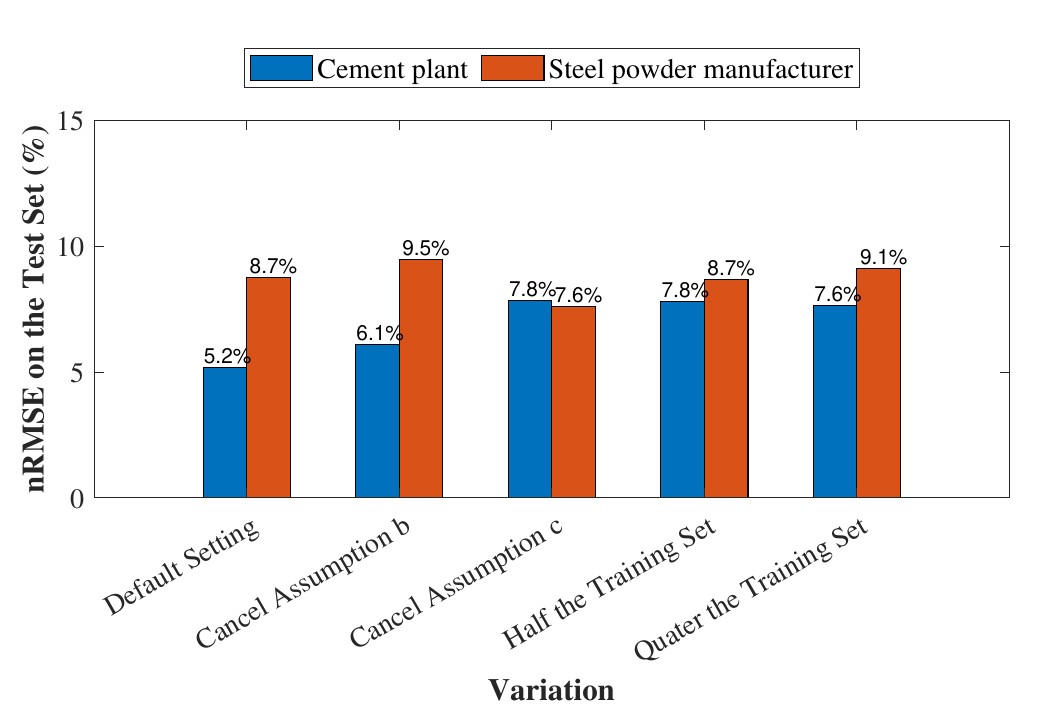}
\caption{Performance of the PSI method with variations.}
  \label{fig_variation}
\end{figure}

\textbf{Setting variations}. Fig.~\ref{fig_variation} shows the performance of the identified load models for the two plants with setting variations, namely, removing prior assumptions b and c described in Section~\ref{sec_training}-B and reducing the number of samples in the training set to $n = 10$ and $n = 5$, respectively. These variations result in reduced PSI performance, which indicates that adding our prior assumption to the model identification problem and increasing the size of the training set help improve the performance of the PSI method. Among these changes, prior assumption b had the largest impact, whereas changing the size of the dataset had a smaller impact, which indicates that the PSI approach is less dependent on the dataset than other methods. When Assumption d does not hold, the load modeling accuracy in the tests considerably decreases, and practical accuracy is not achieved. Therefore, we do not present these results here. Nevertheless, as discussed in Section~\ref{sec_assumption}, Assumption d involves only some prior knowledge and does not involve the private information of industrial users, so it generally holds.

\section{Discussion}\label{sec_discussion}

Although we designed laboratory-level tests to validate the principles of PSI, our method requires notable improvements before it can be applied in practical industry applications. Naturally, this gap arises from the assumptions considered in PSI, which may deviate from actual conditions. Therefore, in this section, we discuss the degree of alignment between the main assumptions of PSI and real conditions, aiming to gradually reduce the reliance on these assumptions through further in-depth research.

First, in our model and numerical tests, we assume that industrial users' energy costs can be calculated simply by multiplying hourly energy consumption with electricity prices. The electricity prices could be day-ahead/real-time prices in the wholesale market or time-of-use prices set by retailers or grid operators.
However, in reality, many industrial users do not directly face fluctuating electricity prices but settle with counterparts according to the terms specified in power purchase agreements with retailers or directly with generators. For example, if the electricity prices are based on long-term contracts or time-of-use prices, which could have long updating cycles (e.g., monthly), then there is less available information to learn from historical data.

Second, we assume that users aim to minimize energy costs while meeting planned production targets, which is the most basic scheduling objective for industrial production in the literature. However, industrial users may have other objectives, such as risk aversion regarding changing electricity costs or minimizing deviations from original production plans.

Finally, we assume that the energy consumption mechanisms of industrial users to be modeled can be effectively captured by the STN model, enabling the construction of an inverse optimization problem based on the mSTN. However, as mentioned earlier, the STN is a general-purpose model that neglects some modeling details and cannot capture features such as parameter uncertainty and the nonlinear production efficiency of industrial loads.

For the aforementioned issues and other potential challenges not addressed in detail in this paper, a general approach is to consider that the PSI framework is theoretically compatible with different load models. Depending on specific requirements and considerations, different model assumptions can be established to modify the default mSTN, with corresponding changes made to other relevant parts of the framework. Given sufficient computational resources, a library can be constructed comprising major industrial load models and user objectives, and the required models can be determined through model training and cross-validation. Nevertheless, further research is needed to explore how to modify the PSI framework to accommodate situations in which our assumption does not hold and to investigate the performance of PSI under such conditions. In future research, we plan to extend the PSI method to include additional realistic considerations, such as nonshiftable loads, the irrationality of users, and production scheduling in the long term. In addition, we plan to test the PSI method on more realistic datasets and develop algorithms to accelerate the model identification process.

\section{Conclusion}\label{sec_conclusion}

In this paper, we present PSI, an inverse-optimization-based framework for modeling the loads of industrial users in cases with incomplete information. In PSI, the load model used to describe the energy-consuming behavior of an industrial user given certain boundary conditions is the modified STN model, and the model parameters are fitted via historical smart meter data.
As shown in the numerical results, PSI was used to effectively determine the parameters of load models of a steel powder manufacturer and a cement plant, with satisfactory accuracy achieved with 21 days of price-consumption data. The modeling errors of the established load models did not exceed 8.5\% and 5.25\%, respectively, compared with the verified-accurate load model obtained from directly accessing private data. The proposed approach achieved significantly higher accuracy than the mainstream pure data-driven approaches.

PSI facilitates effective load modeling of industrial users even though limited coarse-scale external meter data are available. Our work can help address the issue of grid-load interactions where load aggregators or virtual power plants, despite possessing sophisticated industrial load models, are unable to directly obtain model parameters due to privacy concerns.



\section*{Acknowledgments}
Generative AI was used to enhance the grammar and readability of this paper.

\appendices
\section{Supplementary Discussion of Per-Unit Values in the mSTN}\label{app_per_unitization}

Without loss of generality, the state change $\Delta S_i$ of intermediate $i$ (\ref{primal_constraint_changeofS2}) can be expressed as:
\begin{equation}
\Delta S_i = \Delta E_i  g_i - \Delta E_{i + 1}  c_{i + 1}
\end{equation}
where $\Delta E_i = P_i  \Delta t$ is the energy consumed in task $i$ and $g_i(c_{i +1})$ is the production (consumption) of intermediate $i$ associated with consuming one unit of energy in task $i(i+1)$.
Now, we consider the conversion from intermediate $i$ to $i+1$: assuming that only task $i+1$ is implemented and it consumes $\Delta E_{i +1}$ energy, the consumed intermediate $i$ is $\Delta E_{i + 1}  c_{i + 1}$, and the produced intermediate $i + 1$ is $\Delta E_{i + 1}  g_{i + 1}$. Therefore, the state conversion rate of task $i+1$ (state $i$ to $i+1$) is $r_{i, i+1} = \frac{\Delta E_{i + 1}  g_{i + 1}}{\Delta E_{i + 1}  c_{i + 1}} = \frac{g_{i + 1}}{c_{i + 1}}$. Recursively, the conversion rate from state $i$ to $i^{\rm end}$ is
\begin{equation}
  r_{i, i^{\rm end}} = \frac{\Delta S_{i^{\rm end}}}{\Delta S_{i}} =\prod_{j=i+1}^{i^{\rm end}} \frac{g_j}{c_j}
\end{equation}
The per-unit values of $g_i, c_{i+1}, S^{\rm 0/max/tar}_i$ are given on the basis of the obtained values and multiplied by $r_{i, i^{\rm end}}$.
When the per-unit value is used, $g_i = c_i, \forall i$; therefore, we use $g_i$ to denote it in the mSTN.

\section{Supplementary Discussion of Task Aggregation}\label{app_aggregating}
The aggregation strategy is designed to reduce the number of parameters in the mSTN model and enhance the convergence of the parameter identification process. In this work, we prefer to aggregate production links with similar throughput first because these links are likely to be turned on simultaneously in periods with low electricity prices and turned off simultaneously in periods with high electricity prices. Intuitively, the model error caused by this aggregation process is small. The aggregation of tasks $i$ and $i + 1$ is used as an example. It is assumed that $\Delta S_{i - 1}$ of intermediate $i - 1$ is processed in task $i$ and then task $i + 1$ to produce $\Delta S_{i + 1}$ of intermediate $i + 1$. Using the per-unit value, we have $\Delta S_{i - 1} = \Delta S_{i + 1} = \Delta S$. The net electricity consumption of the two tasks is $\Delta E = \sum_{j = i}^{i + 1} \frac{\Delta S}{g_j}$. Let
\begin{equation}
  g_{i'} = (\sum_{j = i}^{i + 1} \frac{1}{g_j})^{-1}, P^{\rm max}_{i'} = \sum_{j = i}^{i + 1} P^{\rm max}_{j}
\end{equation}
and a new task $i'$ parameterized by $g_{i'}, P^{\rm max}_{i'}$ can approximate both tasks $i, i+1$, which neglects the buffer between the two tasks. The above process can be generalized to the aggregation of multiple tasks.

Conceptually, when estimating the parameters of machines before a facility is built, the throughputs of the tasks are often similar, i.e., $g_{i}  P^{\rm max}_{i} \approx g_{i+1}  P^{\rm max}_{i+1}$. This means that the two tasks can simultaneously be performed at rated power when electricity prices are low without accumulating intermediates. Intuitively, adjacent tasks can be aggregated without yielding substantial error.

In the numerical test, to obtain the mSTN models for different numbers of tasks, we first assume that $g_i$ in the 10-task model (Table~\ref{tab_parameter}) is known and then aggregate a task with a former task at a given time via the above aggregation process. The aggregation order is (10, 3, 5, 9, 6, 2, 4, 8). In particular, since machines at different operating points may have different $g_i$ values, we used $g_i$ at the rated power as prior knowledge. For example, for the crusher task, $g=15/20=0.75$.

\section{Configurations of the Machine Learning Approaches}\label{app_ml_methods}
\paragraph{MLP}The MLP model consists of a fully connected neural network with two hidden layers. The nodes in each layer are arranged in a 24-48-48-24 structure. The input is the 24-hour energy price, and the output is the 24-hour electricity consumption. The optimizer is Adam with a learning rate of lr=2e-5 and $\beta=(0.9,0.999)$.
\paragraph{LSTM} The LSTM model has the same input and output structures as the MLP. It utilizes 1 LSTM layer to efficiently capture the relationships in time series. The optimizer is Adam with a learning rate of lr=5e-4 and $\beta=(0.9,0.999)$.
\paragraph{SVR} The SVR model uses an SVM algorithm to fit the data for continuous target variables. To predict 24-hour electricity consumption, 24 independent SVR models are trained separately; each has the same 24-hour energy price as the input, and the corresponding one-hour electricity consumption is predicted and output.

The model hyperparameters listed in the appendix result from our extensive tuning. As mentioned earlier, the scenario we focus on has a limited amount of available data. In this context, machine learning models are likely to display underfitting, which is why we ultimately chose models with few parameters.


\ifCLASSOPTIONcaptionsoff
  \newpage
\fi


\bibliographystyle{IEEEtran}
\bibliography{reference}

@article{zhuo_cost_2022,
	title = {Cost increase in the electricity supply to achieve carbon neutrality in {China}},
	volume = {13},
	issn = {2041-1723},
	language = {en},
	number = {1},
	urldate = {2022-09-16},
	journal = {Nat Commun},
	author = {Zhuo, Zhenyu and Du, Ershun and Zhang, Ning and Nielsen, Chris P. and Lu, Xi and Xiao, Jinyu and Wu, Jiawei and Kang, Chongqing},
	month = jun,
	year = {2022},
	pages = {3172},
}

@inproceedings{pantelides1994unified,
  title     = {Unified frameworks for optimal process planning and scheduling},
  booktitle = {In Proceedings on the second conference on foundations of computer aided operations},
  author    = {Pantelides, C. C.},
  month     = jul,
  year      = {1994},
  pages     = {253--274},
}

@inproceedings{lyu_lstn_2023,
	title = {{LSTN}: {A} {Linear} {Model} of {Industrial} {Production} {Process} for {Demand} {Response}},
	shorttitle = {{LSTN}},
	urldate = {2024-02-01},
	booktitle = {2023 {IEEE} {PES} {Innovative} {Smart} {Grid} {Technologies} {Europe} ({ISGT} {EUROPE})},
	author = {Lyu, Ruike and Guo, Hongye and Zheng, Yuanjie and Bai, Yunlong and Chen, Qixin},
	month = oct,
	year = {2023},
	pages = {1--5},
}

@article{tan_data-driven_2023,
	title = {Data-{Driven} {Inverse} {Optimization} for {Modeling} {Intertemporally} {Responsive} {Loads}},
	issn = {1949-3061},
	volume = {14},
month = sep,
	issn = {1949-3053, 1949-3061},
	language = {en},
	number = {5},
	journal = {IEEE Trans. Smart Grid},
	author = {Tan, Zhenfei and Yan, Zheng and Xia, Qing and Wang, Yang},
	year = {2023},
	pages = {4129 - 4132},
}

@article{golmohamadi_demand-side_2022,
	title = {Demand-side management in industrial sector: {A} review of heavy industries},
	volume = {156},
	issn = {13640321},
	shorttitle = {Demand-side management in industrial sector},
	language = {en},
	urldate = {2023-03-31},
	journal = {Renewable and Sustainable Energy Reviews},
	author = {Golmohamadi, Hessam},
	month = mar,
	year = {2022},
	pages = {111963},
}

@article{sun_clustering-based_2019,
	title = {Clustering-{Based} {Residential} {Baseline} {Estimation}: {A} {Probabilistic} {Perspective}},
	volume = {10},
	issn = {1949-3061},
	shorttitle = {Clustering-{Based} {Residential} {Baseline} {Estimation}},
	number = {6},
	journal = {IEEE Trans. Smart Grid	},
	author = {Sun, Mingyang and Wang, Yi and Teng, Fei and Ye, Yujian and Strbac, Goran and Kang, Chongqing},
	year = {2019},
	pages = {6014--6028},
}

@article{li_precision_2022,
	title = {Precision and {Accuracy} {Co}-optimization {Based} {Demand} {Response} {Baseline} {Load} {Estimation} {Using} {Bi}-directional {Data}},
	issn = {1949-3061},
	journal = {IEEE Trans. Smart Grid	},
	author = {Li, Kangping and Wang, Yuxi and Zhang, Ning and Wang, Fei},
	year = {2022},
	pages = {266-76},
volume={14},
  number={1},
}

@article{gong_integrated_2022,
	title = {Integrated scheduling of hot rolling production planning and power demand response considering order constraints and {TOU} price},
	volume = {16},
	issn = {1751-8687, 1751-8695},
	language = {en},
	number = {14},
	urldate = {2022-10-06},
	journal = {IET Generation Trans \& Dist},
	author = {Gong, Feixiang and Chen, Songsong and Tian, Shiming and Qin, Jian and Zhang, Haijing and Sun, Beibei and Yuan, Jindou and Jiang, Linru and Xu, Yuting and Wang, Yong},
	month = jul,
	year = {2022},
	pages = {2840--2851},
}

@article{lu_data-driven_2021,
	title = {Data-driven real-time price-based demand response for industrial facilities energy management},
	volume = {283},
	issn = {0306-2619},
	language = {en},
	urldate = {2022-10-07},
	journal = {Appl. Energy},
	author = {Lu, Renzhi and Bai, Ruichang and Huang, Yuan and Li, Yuting and Jiang, Junhui and Ding, Yuemin},
	month = feb,
	year = {2021},
	pages = {116291},
}

@article{ding_demand_2014,
	title = {A {Demand} {Response} {Energy} {Management} {Scheme} for {Industrial} {Facilities} in {Smart} {Grid}},
	volume = {10},
	issn = {1941-0050},
	number = {4},
	journal = {IEEE Trans. Ind. Informat.},
	author = {Ding, Yue Min and Hong, Seung Ho and Li, Xiao Hui},
	month = nov,
	year = {2014},
	pages = {2257--2269},
}

@article{kondili_general_1993,
	title = {A general algorithm for short-term scheduling of batch operations—{I}. {MILP} formulation},
	volume = {17},
	issn = {00981354},
	language = {en},
	number = {2},
	urldate = {2022-10-09},
	journal = {Comput. Chem. Eng.},
	author = {Kondili, E. and Pantelides, C.C. and Sargent, R.W.H.},
	month = feb,
	year = {1993},
	pages = {211--227},
}

@inproceedings{dehghan-dehnavi_estimating_2020,
	title = {Estimating {Participation} {Abilities} of {Industrial} {Customers} in {Demand} {Response} {Programs}: {A} {Two}-{Level} {Decision}-{Making} {Tree} {Analysis}},
	shorttitle = {Estimating {Participation} {Abilities} of {Industrial} {Customers} in {Demand} {Response} {Programs}},
	booktitle = {2020 {IEEE}/{IAS} 56th {Indust.} and {Commerc.} {Power} {Syst.} {Tech.} {Conf.} ({I}\&{CPS})},
	author = {Dehghan-Dehnavi, Somayeh and Fotuhi-Firuzabad, Mahmud and Moeini-Aghtaie, Moein and Dehghanian, Payman and Wang, Fei},
	month = jun,
	year = {2020},
	note = {ISSN: 2158-4907},
	pages = {1--8},
}

@inproceedings{zhang_industrial_2015,
	title = {Industrial demand response by steel plants with spinning reserve provision},
	booktitle = {2015 {North} {American} {Power} {Symposium} ({NAPS})},
	author = {Zhang, Xiao and Hug, Gabriela and Kolter, Zico and Harjunkoski, Iiro},
	month = oct,
	year = {2015},
	pages = {1--6},
}

@article{lu_electricity_2022,
	title = {Electricity {Load} {Profile} {Characterisation} for {Industrial} {Users} {Based} on {Normal} {Cloud} {Model} and {iCFSFDP} {Algorithm}},
	journal = {IEEE Trans. Power Syst.},
	author = {Lu, Feng and Cui, Xueyuan and Xing, Jianxu and Liu, Shengyuan and Lin, Zhenzhi and Fei, Xiaoming and Ma, Liang and Huang, Xiang and Ding, Yi and Yang, Li},
	year={2023},
  volume={38},
  number={4},
  pages={3799-3813},
}

@article{zhang_demand_2018,
	title = {Demand {Response} of {Ancillary} {Service} {From} {Industrial} {Loads} {Coordinated} {With} {Energy} {Storage}},
	volume = {33},
	issn = {0885-8950, 1558-0679},
	language = {en},
	number = {1},
	urldate = {2022-10-30},
	journal = {IEEE Trans. Power Syst.},
	author = {Zhang, Xiao and Hug, Gabriela and Kolter, J. Zico and Harjunkoski, Iiro},
	month = jan,
	year = {2018},
	pages = {951--961},
}

@article{lu_multi-agent_2020,
	title = {Multi-agent deep reinforcement learning based demand response for discrete manufacturing systems energy management},
	volume = {276},
	issn = {0306-2619},
	language = {en},
	urldate = {2022-10-31},
	journal = {Appl. Energy},
	author = {Lu, Renzhi and Li, Yi-Chang and Li, Yuting and Jiang, Junhui and Ding, Yuemin},
	month = oct,
	year = {2020},
	pages = {115473},
}

@article{yu_real-time_2016,
	title = {A real-time decision model for industrial load management in a smart grid},
	volume = {183},
	issn = {0306-2619},
	language = {en},
	urldate = {2022-11-09},
	journal = {Appl. Energy},
	author = {Yu, Mengmeng and Lu, Renzhi and Hong, Seung Ho},
	month = dec,
	year = {2016},
	pages = {1488--1497},
}

@article{wohlfarth_demand_2020,
	title = {Demand response in the service sector – {Theoretical}, technical and practical potentials},
	volume = {258},
	issn = {03062619},
	language = {en},
	urldate = {2022-12-25},
	journal = {Appl. Energy},
	author = {Wohlfarth, Katharina and Klobasa, Marian and Gutknecht, Ralph},
	month = jan,
	year = {2020},
	pages = {114089},

}

@article{wang_intelligent_2019,
	title = {Intelligent {Demand} {Response} for {Industrial} {Energy} {Management} {Considering} {Thermostatically} {Controlled} {Loads} and {EVs}},
	volume = {15},
	issn = {1941-0050},
	number = {6},
	journal = {IEEE Trans. Ind. Inf.	},
	author = {Wang, Jidong and Shi, Yingchen and Zhou, Yue},
	month = jun,
	year = {2019},
	pages = {3432--3442},
}

@article{wang_modelling_2019,
	title = {Modelling deep decarbonization of industrial energy consumption under 2-degree target: {Comparing} {China}, {India} and {Western} {Europe}},
	volume = {238},
	issn = {03062619},
	shorttitle = {Modelling deep decarbonization of industrial energy consumption under 2-degree target},
	language = {en},
	urldate = {2022-12-25},
	journal = {Appl. Energy},
	author = {Wang, Huan and Chen, Wenying},
	month = mar,
	year = {2019},
	pages = {1563--1572},
}

@article{si_electric_2021,
	title = {Electric {Load} {Clustering} in {Smart} {Grid}: {Methodologies}, {Applications}, and {Future} {Trends}},
	volume = {9},
	issn = {2196-5420},
	shorttitle = {Electric {Load} {Clustering} in {Smart} {Grid}},
	number = {2},
	journal = {J. Mod. Power Syst. Clean Energy},
	author = {Si, Caomingzhe and Xu, Shenglan and Wan, Can and Chen, Dawei and Cui, Wenkang and Zhao, Junhua},
	month = mar,
	year = {2021},
	pages = {237--252},
}

@article{wang_load_2015,
	title = {Load profiling and its application to demand response: {A} review},
	volume = {20},
	issn = {1007-0214},
	shorttitle = {Load profiling and its application to demand response},
	number = {2},
	journal = {Tsinghua Sci. Technol.	},
	author = {Wang, Yi and Chen, Qixin and Kang, Chongqing and Zhang, Mingming and Wang, Ke and Zhao, Yun},
	month = apr,
	year = {2015},
	pages = {117--129},
}

@article{ahuja_inverse_2001,
	title = {Inverse {Optimization}},
	volume = {49},
	issn = {0030-364X, 1526-5463},
	language = {en},
	number = {5},
	urldate = {2022-12-27},
	journal = {Oper. Res.},
	author = {Ahuja, Ravindra K. and Orlin, James B.},
	month = oct,
	year = {2001},
	pages = {771--783},
}

@inproceedings{Lofberg2004,
address = {Taipei, Taiwan},
author = {L{\"{o}}fberg, J.},
booktitle = {In Proceedings of the CACSD Conference},
title = {YALMIP : A Toolbox for Modeling and Optimization in MATLAB},
year = {2004}
}

@article{cortez_demand_2023,
	title = {Demand {Management} for {Peak} to {Average} {Ratio} {Minimization} via {Intraday} {Block} {Pricing}},
	issn = {1949-3053, 1949-3061},
	doi = {10.1109/TSG.2023.3240522},
	language = {en},
	urldate = {2023-02-09},
	journal = {IEEE Trans. Smart Grid},
	author = {Cortez, Carolina and Kasis, Andreas and Papadaskalopoulos, Dimitrios and Timotheou, Stelios},
	year={2023},
  volume={14},
  number={5},
  pages={3584-3599}
}

@article{tao_customer-centered_2023,
	title = {Customer-{Centered} {Pricing} {Strategy} {Based} on {Privacy}-{Preserving} {Load} {Disaggregation}},
	issn = {1949-3053, 1949-3061},
	doi = {10.1109/TSG.2023.3238029},
	language = {en},
	urldate = {2023-02-09},
	journal = {IEEE Trans. Smart Grid},
	author = {Tao, Yuechuan and Qiu, Jing and Lai, Shuying and Sun, Xianzhuo and Ma, Yuan and Zhao, Junhua},
	year={2023},
  volume={14},
  number={5},
  pages={3401-3412},
}

@article{de_sa_ferreira_time--use_2013,
	title = {Time-of-{Use} {Tariff} {Design} {Under} {Uncertainty} in {Price}-{Elasticities} of {Electricity} {Demand}: {A} {Stochastic} {Optimization} {Approach}},
	volume = {4},
	issn = {1949-3061},
	shorttitle = {Time-of-{Use} {Tariff} {Design} {Under} {Uncertainty} in {Price}-{Elasticities} of {Electricity} {Demand}},
	number = {4},
	journal = {IEEE Trans. Smart Grid	},
	author = {de Sá Ferreira, Rafael and Barroso, Luiz Augusto and Lino, Priscila Rochinha and Carvalho, Martha Martins and Valenzuela, Paula},
	month = dec,
	year = {2013},
	pages = {2285--2295},

}

@article{nojavan_optimal_2017,
	title = {Optimal stochastic energy management of retailer based on selling price determination under smart grid environment in the presence of demand response program},
	volume = {187},
	issn = {03062619},
	language = {en},
	urldate = {2023-02-15},
	journal = {Appl. Energy},
	author = {Nojavan, Sayyad and Zare, Kazem and Mohammadi-Ivatloo, Behnam},
	month = feb,
	year = {2017},
	pages = {449--464},

}

@article{maharjan_dependable_2013,
	title = {Dependable {Demand} {Response} {Management} in the {Smart} {Grid}: {A} {Stackelberg} {Game} {Approach}},
	volume = {4},
	issn = {1949-3061},
	shorttitle = {Dependable {Demand} {Response} {Management} in the {Smart} {Grid}},
	doi = {10.1109/TSG.2012.2223766},
	number = {1},
	journal = {IEEE Trans. Smart Grid	},
	author = {Maharjan, Sabita and Zhu, Quanyan and Zhang, Yan and Gjessing, Stein and Basar, Tamer},
	month = mar,
	year = {2013},
	pages = {120--132},

}

@article{lu_2018_data,
  title={A data-driven Stackelberg market strategy for demand response-enabled distribution systems},
  author={Lu, Tianguang and Wang, Zhaoyu and Wang, Jianhui and Ai, Qian and Wang, Chong},
  journal={IEEE Trans. Smart Grid},
  volume={10},
  number={3},
  pages={2345--2357},
  year={2018},
  publisher={IEEE}
}

@article{chen_2019_learning,
  title={Learning from past bids to participate strategically in day-ahead electricity markets},
  author={Chen, Ruidi and Paschalidis, Ioannis Ch and Caramanis, Michael C and Andrianesis, Panagiotis},
  journal={IEEE Trans. Smart Grid},
  volume={10},
  number={5},
  pages={5794--5806},
  year={2019},
  publisher={IEEE}
}

@article{ruiz_2013_revealing,
  title={Revealing rival marginal offer prices via inverse optimization},
  author={Ruiz, Carlos and Conejo, Antonio J and Bertsimas, Dimitris J},
  journal={IEEE Trans. Power Syst.},
  volume={28},
  number={3},
  pages={3056--3064},
  year={2013},
  publisher={IEEE}
}

@article{li_mixed-integer_2023,
	title = {A mixed-integer programming approach for industrial non-intrusive load monitoring},
	volume = {330},
	issn = {03062619},
	language = {en},
	urldate = {2024-05-18},
	journal = {Appl. Energy},
	author = {Li, Chuyi and Zheng, Kedi and Guo, Hongye and Chen, Qixin},
	month = jan,
	year = {2023},
	pages = {120295},
}

@article{castro_resourcetask_2013,
	title = {Resource–{Task} {Network} {Formulations} for {Industrial} {Demand} {Side} {Management} of a {Steel} {Plant}},
	volume = {52},
	issn = {0888-5885, 1520-5045},
	language = {en},
	number = {36},
	urldate = {2023-03-14},
	journal = {Industrial \& Engineering Chemistry Research},
	author = {Castro, Pedro M. and Sun, Lige and Harjunkoski, Iiro},
	month = sep,
	year = {2013},
	pages = {13046--13058},
}

@article{zhang_cost-effective_2017,
	title = {Cost-{Effective} {Scheduling} of {Steel} {Plants} {With} {Flexible} {EAFs}},
	volume = {8},
	issn = {1949-3053, 1949-3061},
	language = {en},
	number = {1},
	urldate = {2023-03-13},
	journal = {IEEE Trans. Smart Grid},
	author = {Zhang, Xiao and Hug, Gabriela and Harjunkoski, Iiro},
	month = jan,
	year = {2017},
	pages = {239--249},
}

@article{li_real-time_2017,
	title = {Real-{Time} {Demand} {Bidding} for {Energy} {Management} in {Discrete} {Manufacturing} {Facilities}},
	volume = {64},
	issn = {1557-9948},
	doi = {10.1109/TIE.2016.2599479},
	number = {1},
	journal = {IEEE Trans. Ind. Electron.},
	author = {Li, Yi-Chang and Hong, Seung Ho},
	month = jan,
	year = {2017},
	pages = {739--749},
}

@article{zhang_data-driven_2023,
	title = {Data-{Driven} {Security} and {Stability} {Rule} in {High} {Renewable} {Penetrated} {Power} {System} {Operation}},
	volume = {111},
	issn = {1558-2256},
	doi = {10.1109/JPROC.2022.3192719},
	number = {7},
	journal = {Proc. IEEE},
	author = {Zhang, Ning and Jia, Hongyang and Hou, Qingchun and Zhang, Ziyang and Xia, Tian and Cai, Xiao and Wang, Jiaxin},
	month = jul,
	year = {2023},
	pages = {788--805},
}

@article{chen_real-time_2024,
	title = {Real-time operation strategy of virtual power plants with optimal power disaggregation among heterogeneous resources},
	volume = {361},
	issn = {0306-2619},
	urldate = {2024-02-25},
	journal = {Appl. Energy},
	author = {Chen, Qixin and Lyu, Ruike and Guo, Hongye and Su, Xiangbo},
	month = may,
	year = {2024},
	pages = {122876},
}

@article{golmohamadi_robust_2020,
	title = {Robust {Self}-{Scheduling} of {Operational} {Processes} for {Industrial} {Demand} {Response} {Aggregators}},
	volume = {67},
	language = {en},
	number = {2},
	journal = { IEEE Trans. Ind. Electron.},
	author = {Golmohamadi, Hessam and Keypour, Reza and Bak-Jensen, Birgitte and Pillai, Jayakrishnan R and Khooban, Mohammad Hassan},
	year = {2020},
}

@misc{rick10119_psi_2024,
	author = {Ruike Lyu},
	url = {https://github.com/Rick10119/Production-Scheduling-Identification},
	urldate = {2024-09-07},
}

@misc{liu_primer_2020,
	title = {A {Primer} on {Zeroth}-{Order} {Optimization} in {Signal} {Processing} and {Machine} {Learning}},
	url = {http://arxiv.org/abs/2006.06224},
	language = {en},
	urldate = {2024-08-21},
	publisher = {arXiv},
	author = {Liu, Sijia and Chen, Pin-Yu and Kailkhura, Bhavya and Zhang, Gaoyuan and Hero, Alfred and Varshney, Pramod K.},
	month = jun,
	year = {2020},
	note = {arXiv:2006.06224 [cs, eess, stat]},
}

@misc{duck_curve,
	title = {California’s electricity duck curve is deepening},
	url = {https://www.pv-magazine.com/2023/07/06/californias-electricity-duck-curve-is-deepening/},
	language = {en-US},
	urldate = {2024-08-26},
	journal = {pv magazine International},
	month = jul,
	year = {2023},
}

@misc{noauthor_-1h24____nodate,
	title = {Shandong Electricity Market {1H24 Review}},
	url = {https://stock.finance.sina.com.cn/stock/go.php/vReport_Show/kind/strategy/rptid/774984472737/index.phtml},
	urldate = {2024-08-26},
}

@article{niu_enhanced_2023,
	title = {Enhanced flexibility utilization and coordinated dispatch method of energy‐intensive enterprises in power systems under time of use prices},
	volume = {17},
	issn = {1752-1416, 1752-1424},
	language = {en},
	number = {15},
	urldate = {2024-08-26},
	journal = {IET Renew. Power Gener.},
	author = {Niu, Tao and Li, Fan and Fang, Sidun},
	month = nov,
	year = {2023},
	pages = {3609--3623},
}

@article{chen_pathway_2021,
	title = {Pathway toward carbon-neutral electrical systems in {China} by mid-century with negative {CO2} abatement costs informed by high-resolution modeling},
	volume = {5},
	issn = {25424351},
	language = {en},
	number = {10},
	urldate = {2022-09-19},
	journal = {Joule},
	author = {Chen, Xinyu and Liu, Yaxing and Wang, Qin and Lv, Jiajun and Wen, Jinyu and Chen, Xia and Kang, Chongqing and Cheng, Shijie and McElroy, Michael B.},
	month = oct,
	year = {2021},
	pages = {2715--2741},
}

@article{wang_review_2019,
	title = {Review of {Smart} {Meter} {Data} {Analytics}: {Applications}, {Methodologies}, and {Challenges}},
	volume = {10},
	copyright = {https://ieeexplore.ieee.org/Xplorehelp/downloads/license-information/IEEE.html},
	issn = {1949-3053, 1949-3061},
	shorttitle = {Review of {Smart} {Meter} {Data} {Analytics}},
	language = {en},
	number = {3},
	urldate = {2024-08-27},
	journal = {IEEE Trans. Smart Grid},
	author = {Wang, Yi and Chen, Qixin and Hong, Tao and Kang, Chongqing},
	month = may,
	year = {2019},
	pages = {3125--3148},
}

\begin{IEEEbiography}[{\includegraphics[width=1.1in,height=1.5in,clip,keepaspectratio]{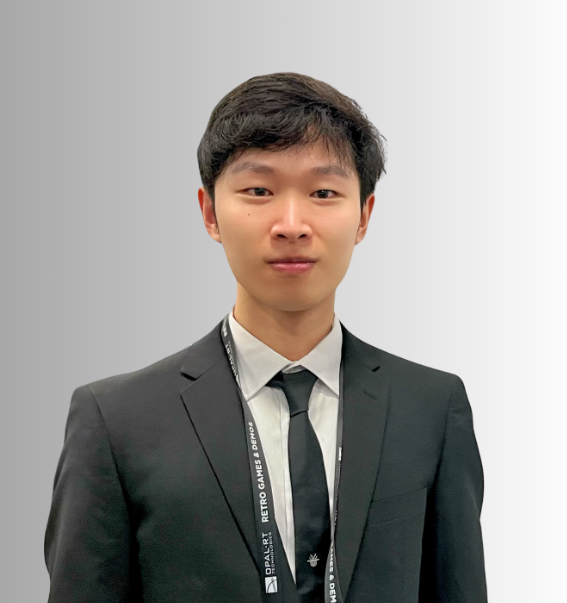}}]{Ruike Lyu}

  received the bachelor's degree in electrical engineering from Tsinghua University, Beijing, China, in 2021, where he is currently pursuing the Ph.D. degree. His research focuses on the interaction between industrial load and the power grid.
  
\end{IEEEbiography}

\begin{IEEEbiography}[{\includegraphics[width=1in,height=1.25in,clip,keepaspectratio]{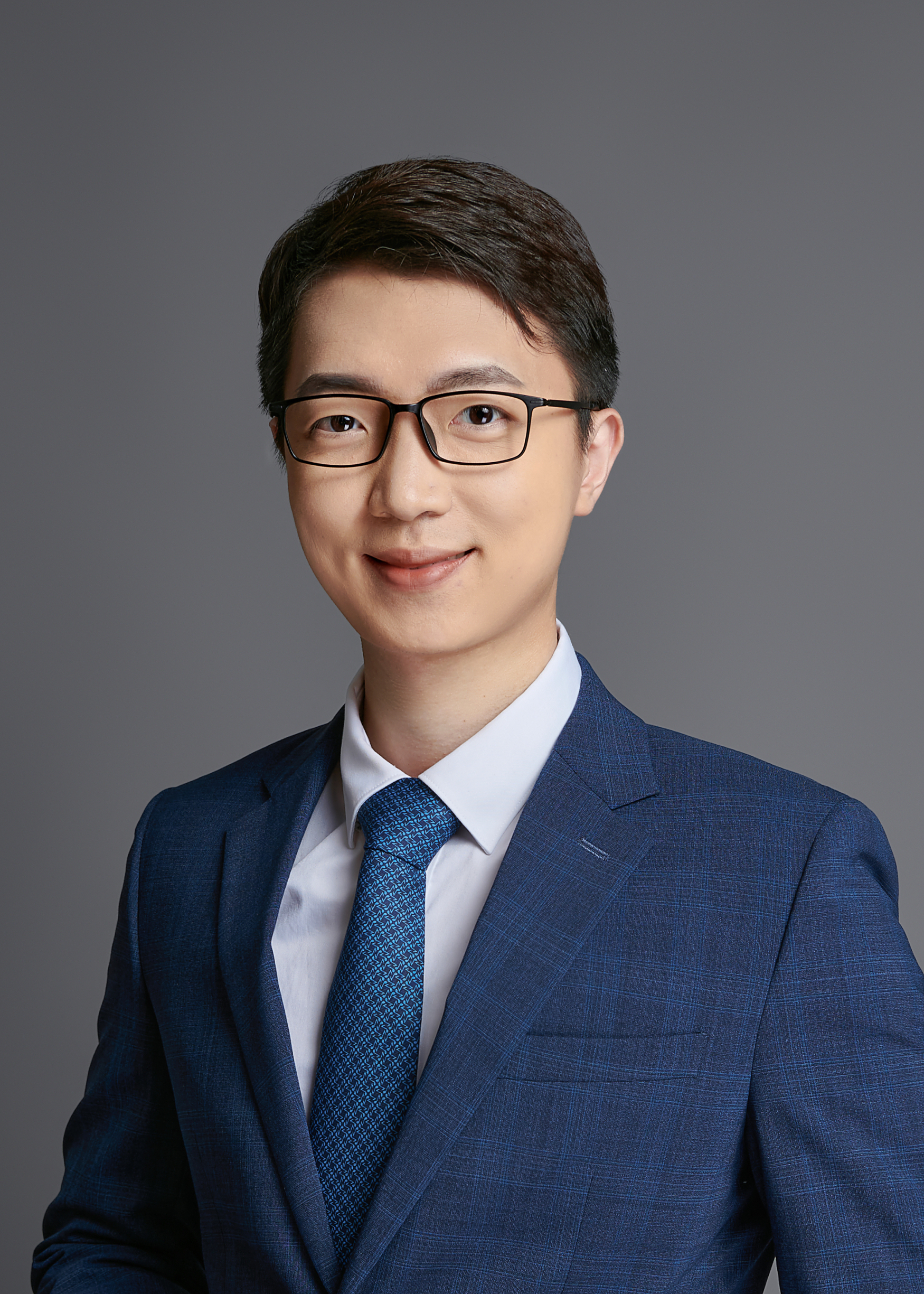}}]{Hongye Guo}

  (S'15-M'20) received the B.S. and Ph.D. degrees in electrical engineering from Tsinghua University, Beijing, China, in 2015 and 2020, respectively. He was a post-doctor in electrical engineering from Tsinghua University from 2020 to 2022. He was a visiting student researcher with Stanford University, CA, USA, in 2018, and with Illinois Institute of Technology, Chicago, IL, USA, in 2019. He is currently an assistant professor at Tsinghua University. His research interests include electricity markets, demand-side flexibility, and machine learning applications in power markets.
  
\end{IEEEbiography}

\begin{IEEEbiography}[{\includegraphics[width=1in,height=1.25in,clip,keepaspectratio]{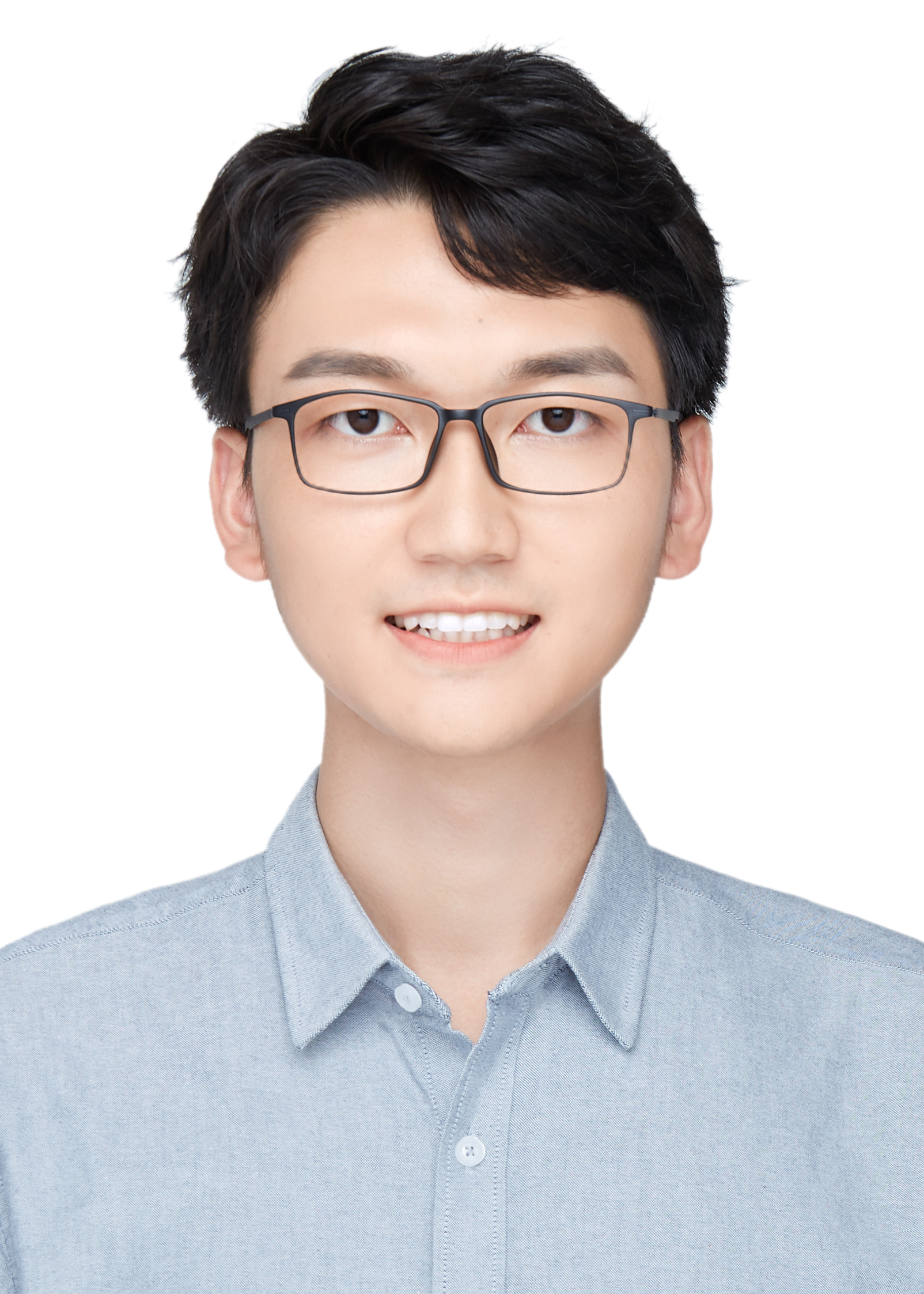}}]{Qinghu Tang}

received the B.S. degree in 2020 from the Department of Electrical Engineering, Tsinghua University, Beijing, China, where he is currently working toward the Ph.D. degree. His research interests include electricity markets, data-driven analysis, and application of machine learning.
  
\end{IEEEbiography}

\begin{IEEEbiography}[{\includegraphics[width=1.0in,height=1.25in,clip,keepaspectratio]{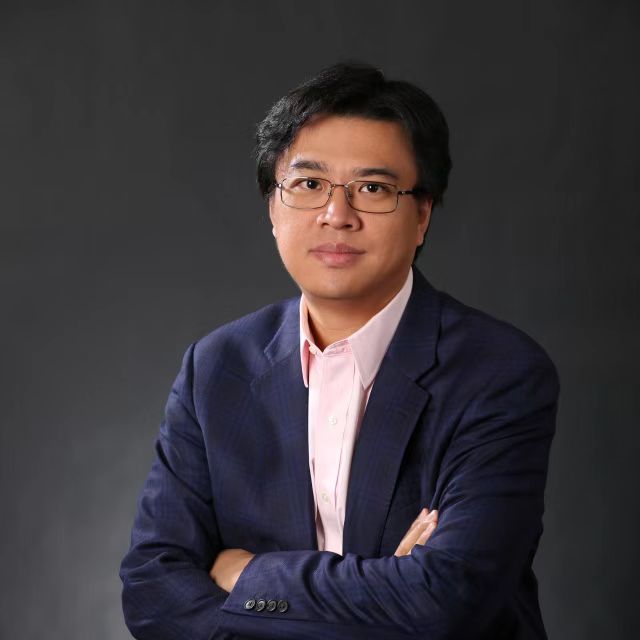}}]{Qixin Chen}
(Senior Member, IEEE) received the Ph.D. degree from the Department of Electrical Engineering, Tsinghua University, Beijing, China, in 2010. His research interests include electricity markets, power system economics and optimization, low-carbon electricity, and power generation expansion planning.
\end{IEEEbiography}

\begin{IEEEbiography}[{\includegraphics[width=1in,height=1.25in,clip,keepaspectratio]{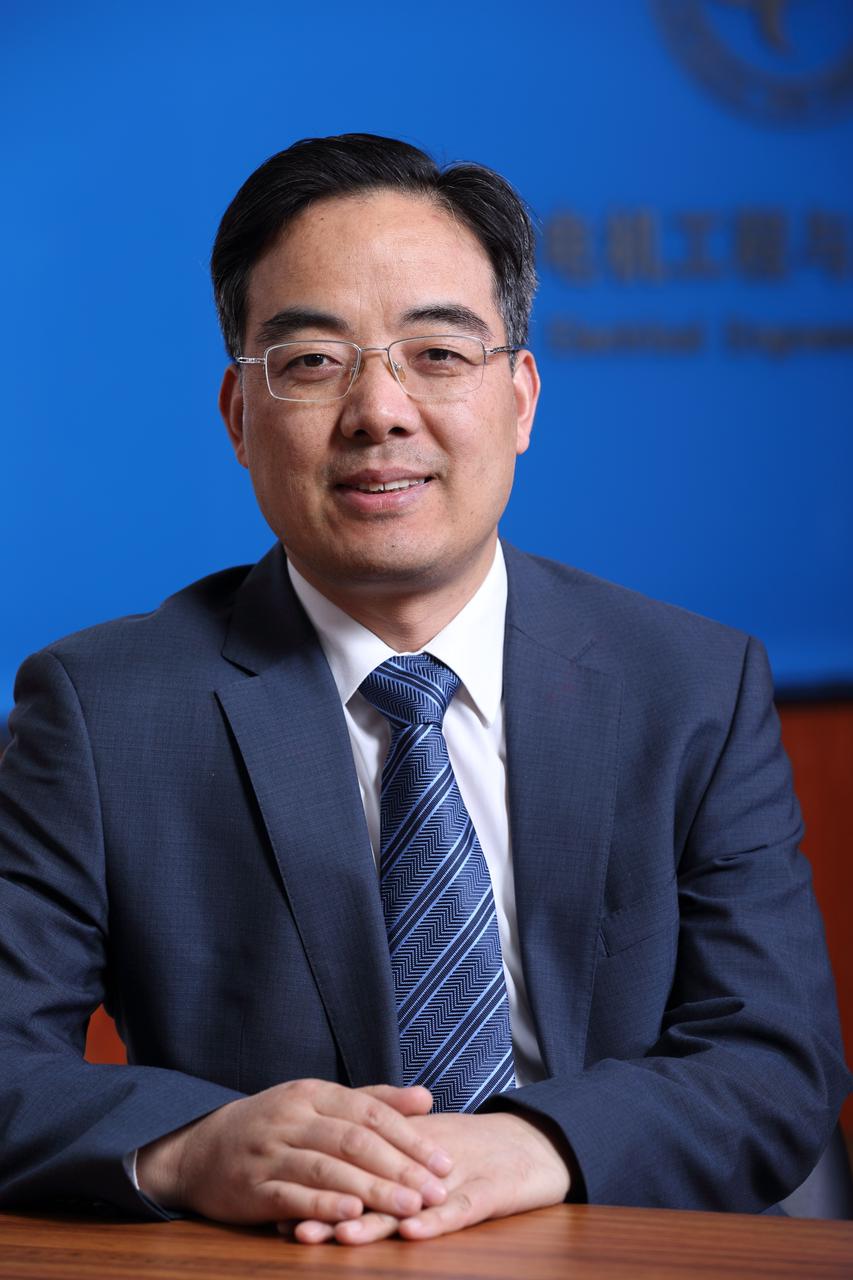}}]{Chongqing Kang}

 (Fellow, IEEE) received the
 Ph.D. degree from the Department of Electrical
 Engineering, Tsinghua University, Beijing, China,
 in 1997, where he is currently a Professor. His
 research interests include power system planning,
 power system operation, renewable energy, low-carbon electricity technology, and load forecasting.
  
\end{IEEEbiography}

\end{document}